\documentclass[journal, 10pt, letterpaper]{IEEEtran}
\usepackage[T1]{fontenc}
\usepackage{url}
\ifCLASSINFOpdf
\else
\fi

\usepackage{amsmath}
\usepackage{algorithmicx}
\usepackage{algorithm}
\usepackage{graphicx}

\usepackage{amssymb}
\usepackage{algpseudocode}
\usepackage{nccmath,textcomp}
\usepackage{eso-pic}
\newcommand{\Mod}[1]{\ (\text{mod}\ #1)}
\makeatletter
\let\MYcaption\@makecaption
\makeatother

\usepackage[font=footnotesize]{subcaption}

\makeatletter
\let\@makecaption\MYcaption
\makeatother
\begin{document}
\title{Adaptive Pilot Patterns for CA-OFDM Systems in Nonstationary Wireless Channels}
\author{Raghunandan M. Rao,~\IEEEmembership{Student Member,~IEEE},
        Vuk Marojevic,~\IEEEmembership{Member,~IEEE},
        and~Jeffrey H. Reed,~\IEEEmembership{Fellow,~IEEE}
\thanks{This is the author's version of the work. Personal use of this material is permitted. For citation purposes, the definitive version of record of this work is: R. M. Rao, V. Marojevic and J. H. Reed, ``Adaptive Pilot Patterns for CA-OFDM Systems in Nonstationary Wireless Channels'', To appear in IEEE Transactions of Vehicular Technology, 2017}
\thanks{Manuscript received March 2, 2017; revised July 22,2017; accepted September 3, 2017. Raghunandan M. Rao, Vuk Marojevic and Jeffrey H. Reed were supported by the National Science Foundation (NSF) under Grant CNS-1642873. The work of Jeffrey H. Reed was partly supported by NSF under Grant CNS-1564148.}
\thanks{Raghunandan M. Rao, Vuk Marojevic and Jeffrey H. Reed are with the Bradley Department of Electrical and Computer Engineering, Virginia Tech, Blacksburg, VA, 24060 USA (e-mail: raghumr@vt.edu; maroje@vt.edu; reedjh@vt.edu)}%
}

\markboth{IEEE Transactions on Vehicular Technology, TO APPEAR}%
{Shell \MakeLowercase{\textit{et al.}}: Bare Demo of IEEEtran.cls for IEEE Communications Society Journals}

\maketitle
\begin{abstract}
In this paper, we investigate the performance gains of adapting pilot spacing and power for Carrier Aggregation (CA)-OFDM systems in nonstationary wireless channels. In current multi-band CA-OFDM wireless networks, all component carriers use the same pilot density, which is designed for poor channel environments. This leads to unnecessary pilot overhead in good channel conditions and performance degradation in the worst channel conditions. We propose adaptation of pilot spacing and power using a codebook-based approach, where the transmitter and receiver exchange information about the fading characteristics of the channel over a short period of time, which are stored as entries in a channel profile codebook. We present a heuristic algorithm that maximizes the achievable rate by finding the optimal pilot spacing and power, from a set of candidate pilot configurations. We also analyze the computational complexity of our proposed algorithm and the feedback overhead. We describe methods to minimize the computation and feedback requirements for our algorithm in multi-band CA scenarios and present simulation results in typical terrestrial and air-to-ground/air-to-air nonstationary channels. Our results show that significant performance gains can be achieved when adopting adaptive pilot spacing and power allocation in nonstationary channels. We also discuss important practical considerations and provide guidelines to implement adaptive pilot spacing in CA-OFDM systems. 
\end{abstract}


\IEEEpeerreviewmaketitle
\section{Introduction}
\IEEEPARstart{T}{he} design of fifth generation (5G) wireless networks are currently being investigated \cite{Agiwal_5G_Comp_Survey} and the Third Generation Partnership Project (3GPP) is targeting the freeze of the first release of 5G specifications, Release 15, in 2018 \cite{3GPP_Rel15}. Compared to the current 4G wireless networks, 5G is proposed to bring performance enhancements in capacity, latency, coverage, spectrum utilization, and the ability to handle heterogeneous traffic \cite{Agiwal_5G_Comp_Survey}. A capacity enhancement of $1000 \times$ is being targeted to connect of billions of low power/low throughput devices to the internet, and support machine type communications between these devices. At the physical layer, a spectral efficiency enhancement of $10 \times$ is being targeted \cite{NTIA_5G_Subcom}. 
In addition, waveform flexibility will be the key to enhance spectral efficiency while supporting users under different channel conditions such as terrestrial (frequency selective, low/high mobility), air to ground (frequency flat, low/high mobility) or combinations of these two. Spectrum aggregation, while being a part of current 4G standards, is also considered as a potential 5G technology because of its ability to increase the utilization of fragmented spectrum.

In current wireless standards based on Orthogonal Frequency Division Multiplexing (OFDM), there is little flexibility for adaptive signaling, such as support for multiple classes of adaptive waveform parameters such as subcarrier spacing, OFDM symbol duration, frame structures and adaptive control channel overhead based on different operating conditions. In the evolution from 4G to 5G, there is considerable interest in the research community to adopt multicarrier waveforms with adaptive transmission parameters at the physical layer in order to enhance the spectral efficiency \cite{Schwarz_Rupp_commag_2016, Zhu_Comp_CSIT_MassMIMO_2016, Jaber_Vuk_UAS_2017}. Although not all control channels can be eliminated to reduce system overhead, one class of control signals whose overhead can be controlled are `pilots' or `reference signals'. Pilot signals are known to the receiver, which aids in channel estimation, equalization and link adaptation \cite{sesia2011lte}.  Most standards define a fixed number of pilots to be deployed, but it is a waste of resources when the channel remains flat in time or frequency, or both. 

\subsection{Motivation for Adaptive Pilot Configurations}
\begin{figure}[t]
\centering
\includegraphics[width=3.2in]{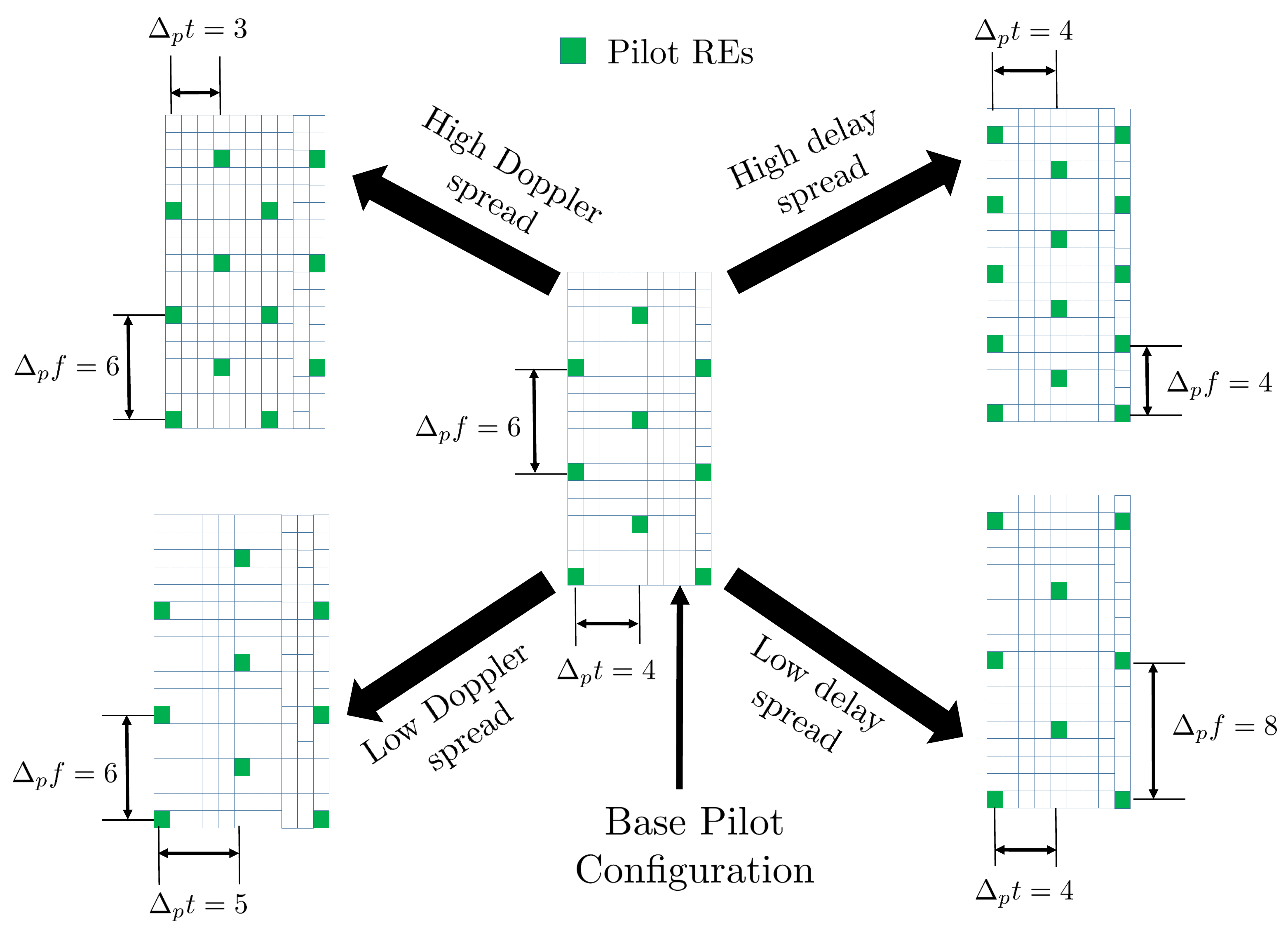}
\caption{Illustration of pilot adaptation in the OFDM resource grid based on varying channel conditions. The figures shows a portion of the time-frequency resource grid. The pilot pattern can be adapted over time, adapting to changing channel statistics at a suitable time granularity.}
\label{Fig1_pilot_adapt_illustrate}
\end{figure}

Wireless channels exhibit different characteristics based on the terrain, propagation environment, obstructions, mobility of users etc. For low mobility and strong line of sight (LoS) channels, the channel is flat in time and frequency, while for high mobility with a strong multipath environment, the channel exhibits strong frequency selectivity and fast temporal fading. Most wireless standards are designed to operate in the worst channel conditions. For this reason the pilot spacing in LTE is designed to satisfactorily capture channel variations for root mean square delay spread $\tau_{rms} = 991 \text{ ns}$ and a user velocity of $500 \text{ km/h}$ at a center frequency $f_c = 2 \text{ GHz}$ \cite{sesia2011lte}. But the wireless channel statistics might be better for a significant number of users at a given point of time. 
The central idea of pilot adaptation is shown in Fig. \ref{Fig1_pilot_adapt_illustrate} where (a) pilot spacing along the time axis is a function of the coherence time of the channel; it is increased when the coherence time is high and decreased when it is low, and (b) pilot spacing along the frequency axis is a function of the coherence bandwidth of the channel; it is increased when the coherence bandwidth of the channel is high and decreased when it is low.

For vehicular-to-vehicular (V2V) and air-to-ground channels, the fading environment can change rapidly and significantly:  
\begin{enumerate}
\item channel temporal correlation varies due to changes in doppler frequency $f_d$ ($f_d$ scales linearly with vehicular velocity). This occurs when the vehicle accelerates, decelerates or changes its direction. 
\item channel spectral correlation varies due to changes in scattering environment as a vehicle moves from one multipath environment to another, and
\item channel spatial correlation varies due to changes in angular spread as a vehicle moves from one scattering environment to another.
\end{enumerate}
Hence, V2V and air-to-ground channels are \emph{nonstationary} and are more likely to benefit from pilot spacing and power adaptation, which is the main focus of this paper.

\subsection{Related Work}
In the past, there has been research in pilot adaptation, where the aim is to vary pilot spacing (also known as pilot periods) and power to meet/maximize a particular target metric with minimal pilot overhead. Since maximizing spectral efficiency is of paramount importance to 5G PHY layer technologies, we consider the metrics fundamentally defined by these objectives such as \emph{capacity, achievable rate, throughput} etc. 

Byun et al. \cite{byun2009adaptive} aim to minimize feedback delays and synchronization mismatch of pilot spacing information in an OFDM system. The authors prioritized maximization of bit error rate (BER) and channel estimation mean square error (MSE), sometimes at the cost of spectral efficiency. Ali et al. \cite{ali2008adaptive} adapt the pilot distribution in OFDM-based WLAN according to the variation level of the channel to maximize the throughput. 
Sheng at al. \cite{sheng2015data} propose to maximize the sum rate using a power allocation scheme between pilot and data symbols for OFDM in a high-speed train (HST) environment. The authors use an information-theoretic approach to solve this problem, by first estimating the average channel complex gains and then using it in a HST basis expansion channel model to formulate a rate-maximization problem. Karami and Beaulieu \cite{karami2012channel} design a joint adaptive power loading and pilot spacing algorithm to maximize the average mutual information between the input and output of OFDM systems. 
Simko, Wang and Rupp \cite{vsimko2012optimal} consider optimal power allocation between pilot and data symbols in an OFDM system, and apply it to a LTE system. The authors consider two channel estimation algorithms: Least Squares (LS) and Linear MMSE (LMMSE). 
Simko et al. \cite{simko2013adaptive} consider joint optimization of pilot spacing and power for SISO and MIMO-OFDM systems (without carrier aggregation). They propose mapping the pilot pattern to the channel quality indicator (CQI) of LTE. 

The idea of pilot parameter adaptation has also been proposed for multi-user MIMO and 5G technologies such as massive MIMO. Kim et. al \cite{Kim_MU_MIMO_PilLink_adapt_2014} proposed an uplink and downlink pilot power and rate adaptation approach to improve energy efficiency. When adapting pilot spacing, Ksairi et. al \cite{Ksairi_5G_MUMIMO_PilAdapt_2016} proposed a scheduling algorithm to group users with similar channel statistics to improve spectral efficiency. Zhu et. al \cite{Zhu_Comp_CSIT_MassMIMO_2016} designed a closed-loop compressive CSIT feedback and estimation framework in sparse multi-user (MU) massive MIMO channels to improve the CSIT estimation performance. They also designed a learning framework to use the minimum pilot and feedback resources needed under unknown and time-varying channel sparsity levels. Adapting pilot density has also been proposed to increase energy efficiency in future green networks, where the pilot density is increased in high traffic scenarios, and decreased in low traffic periods \cite{Xu_He-Zhang_greenLTE_2013}.

\subsection{Contributions}

The key contributions of this paper are:
\begin{enumerate}
\item We derive closed-form expressions for the channel estimation MSE for OFDM pilots arranged in a ``diamond-pattern''. Compared to the state of the art \cite{vsimko2012optimal}, \cite{simko2013adaptive} our expressions (a) are analytical in order to individually isolate the effect of mobility (time fading) and multipath (frequency fading) on channel estimation MSE, and (b) can be used to quickly recompute the MSE for any general multi-band CA-OFDM configuration.


\item We provide a new scheme to adapt pilot patterns in nonstationary channels using feedback of indices from a `channel statistics codebook', with low complexity and feedback overhead. 

\item We extend this framework to multi-band CA-OFDM systems with reduced feedback requirements.

\item We show the gain in the achievable rate using our pilot adaptation algorithm w.r.t. LTE's pilot pattern, by means of numerical simulations. 

\item We quantify the value of pilot adaptation alone, and make the performance comparison agnostic to protocol-specific mechanisms such as adaptive modulation and coding (AMC).
\end{enumerate}
The processing overhead due to our scheme is negligible since we reuse operations which are already present in modern wireless receivers. The feedback overhead is also negligible since we perform pilot adaptation at a longer timescale, as we will discuss later in the paper.

The rest of the paper is organized as follows. 
Section \ref{prob_form} provides the details of the mathematical formulation of the cost function used to find the optimal pilot configuration. Section \ref{ChEst_MSE} outlines the derivation of closed form expressions for the channel estimation MSE. Section \ref{opt_pil_spac_pow} presents the details of our algorithm based on formation of a codebook of channel profiles. Section \ref{num_results} shows the gains in achievable rate using adaptive pilot configurations compared to fixed pilot configurations for CA-OFDM in a variety of nonstationary wireless channel scenarios. We also provide a comparison of our scheme against other schemes. Section \ref{Pract_consid} discusses practical considerations necessary to incorporate adaptive pilot configurations in wireless standards. Finally, section \ref{conc} concludes the paper.

\subsubsection*{Notation}
The notation used in this paper is as follows. $\mathbb{E}[\cdot]$ and $\mathrm{Var}[\cdot]$ stand for the expectation and variance respectively. Symbols in bold such as $\mathbf{X}$ denotes vector/matrix quantities. Hermitian transpose of $\mathbf{X}$ is represented by $\mathbf{X}^H$ and estimated quantities by the hat symbol $\hat{[\ \cdot \ ]}$. $\lceil \cdot \rceil$ and $\lfloor \cdot \rfloor$ indicate the ceiling and floor operations. $x^*,\Re(x),\text{ and } \lvert x \rvert$ denote the conjugate, real part and magnitude of $x$. Sets are indicated by calligraphic letters such as $\mathcal{X}$. In the context of sets, $\lvert \mathcal{X} \rvert$ stands for the cardinality of $\mathcal{X}$. 
Operators $\text{mode}(\mathbf{z})$ and $\| \mathbf{z} \|$ denotes the mode and Euclidean norm of vector $\mathbf{z}$ respectively. The most important parameters are shown in Table \ref{PilAdaptSymAndNot}. 

\begin{table}[t]
\renewcommand{\arraystretch}{1.2}
\caption{Description of the most important parameters}
\label{PilAdaptSymAndNot}
\centering
\begin{tabular}{l l}
\hline
Variable & Description\\
\hline
$\rho$ & The data to pilot power ratio \\
$\sigma_d^2$ & Average power of data symbols \\
$\sigma_p^2$ & Average power of pilot symbols \\
$\Delta_p t$ & Pilot spacing in time \\
$\Delta_p f$ & Pilot spacing in frequency \\
$\bar{\gamma}$ & Post-equalization SINR \\
$\sigma_{ICI}^2$ & Inter-carrier interference power \\
$\sigma_w^2$ & Noise power \\
$\delta_d$ & Channel estimation MSE of data resource elements \\
$N_{tx}$ & Number of transmit antennas \\
$N_{rx}$ & Number of receive antennas \\
$N$ & Number of subcarriers per OFDM symbol \\
$T_{ofdm}$ & Number of OFDM symbols used for channel statistics \\
 & estimation \\
$\hat{\mathbf{H}}$ & The $N \times T_{ofdm}$ channel matrix used to estimate the \\
  & channel spectral and temporal correlation functions\\
$T_s$ & OFDM symbol duration \\
$f_d$ & Maximum Doppler frequency \\
$\tau_{rms}$ & Root mean square delay spread \\
$\mathbf{\hat{R}_t}$ & $(N_{\Delta t }\times 1)$ vector of estimated channel temporal correlation\\
$\mathbf{\hat{R}_f}$ & $(N_{\Delta f} \times 1)$ vector of estimated channel spectral correlation\\ 
\hline
\end{tabular} 
\end{table}
\section{Problem Formulation}\label{prob_form}
There is wide agreement that instantaneous achievable rate is the best indicator of the throughput of a wireless system \cite{simko2013adaptive, wimax_evaluation}. Since it is not possible to know the instantaneous rate beforehand, we maximize the upper bound of the achievable rate based on estimation of necessary operating parameters \cite{simko2013adaptive}. 

It is to be noted that second order statistics such as power spectrum and correlation do not exist for a nonstationary process. However, statistics such as \textit{time-dependent} correlation functions and spectra can be defined for these processes, by means of the \textit{Local Scattering Function} (see \cite{Non_WSSUS_Matz_2005}). Hence, nonstationary channels whose statistics vary in time and frequency can be modeled as locally stationary \cite{Bernado_vehic_nonstat_2014} using this formulation. However, the time scale over which we assume channel stationarity is crucial to accurately model nonstationary channels in a tractable manner. In this regard, the channel measurement results in \cite{Bernado_vehic_nonstat_2014} show that for nonstationary vehicular environments the time dependent doppler and rms delay spreads remain fairly constant for hundreds of milliseconds. Therefore, we assume similar timescales for channel stationarity in this paper. 


Pilot adaptation can be formulated as a maximization problem of the upper bound of the achievable rate \cite{simko2013adaptive, wimax_evaluation}
\begin{align}
\label{optimization_problem}
\underset{\rho, \Delta_p f, \Delta_p t} {\text{maximize}} \quad & S(\Delta_p f, \Delta_p t) \cdot \log_2 (1 + \overline{\gamma})  \\
\text{subject to} \quad
& \overline{P}_t (\rho, \Delta_p f, \Delta_p t ) \leq 1 \nonumber \\
& 1 \leq \Delta_p t \leq T_{max} \nonumber \\
& 2 \leq \Delta_p f \leq F_{max} \text{ and } \Delta_p f \Mod{2} = 0 \nonumber \\
& \rho \leq \rho_{max}, \nonumber
\end{align}
\noindent where $\Delta_p t$ is the pilot spacing in time, $\Delta_p f$ the pilot spacing in frequency and $\rho=\sigma_d^2 / \sigma_p^2$ the data to pilot power ratio. $\sigma_d^2$ is the transmitted power for data symbols and $\sigma_p^2$ the transmitted power per pilot RE. $\overline{\gamma}$ is the post-equalization SINR under imperfect channel knowledge, $S(\Delta_p f, \Delta_p t)$ is the spectrum utilization function as a function of pilot spacing for OFDM, $\overline{P}_t$ is the average power per resource element (RE), and $T_{max}$ is a function of the maximum tolerable latency by the receiver. 
Pilot spacing in the frequency domain is dictated by the sampling theorem. If $\tau_{max}$ is the maximum excess delay of the channel and $T$ the sampling interval, then by sampling theorem \cite{ChanEstOFDM_Param_model} we have
\begin{equation}
\frac{N}{\Delta_p f} > \frac{\tau_{max}}{T}.
\end{equation}
\indent Therefore $\max (\Delta_p f) =  F_{max} = \lceil \frac{NT}{\tau_{max}}\rceil$ is the maximum allowable pilot spacing that is dictated by the maximum excess delay. If we space the channel taps in the Power Delay Profile (PDP) uniformly, then $F_{max}$ depends on the  maximum number of resolvable multipath components $\tau_{max}/T$. Pilots on alternate pilot bearing OFDM symbols are offset by an index of $\Delta_p f /2$ subcarriers, as shown in Fig. \ref{Fig1_pilot_adapt_illustrate}. It has been shown that channel estimation is optimal when the pilots spacing is equal and diamond-shaped \cite{choi2005optimum}. To satisfy this pattern, $\Delta_p f/2$ must be a positive integer. Therefore the additional constraint $\Delta_p f \Mod{2} = 0$ ensures that the $\Delta_p f$ is even and hence, an optimal symmetric `diamond-shaped' pilot pattern can be obtained. $\rho_{max}$ is the maximum allowable data to pilot power ratio, which is dictated by peak to average power ratio (PAPR) considerations and high-power amplifier (HPA) characteristics. In this work, we consider the Zero Forcing (ZF) Receiver, whose post-equalization SINR $\overline{\gamma}$ is given as \cite{simko2013adaptive}
\begin{equation}
\label{posteq_SINR}
\overline{\gamma} = \frac{\sigma_d^2}{\sigma_w^2 + \sigma_{ICI}^2 + \sigma_d^2  \cdot \delta_{d} } \sigma_{ZF},
\end{equation}
\noindent where $\sigma_w^2$ is the average noise power and $\delta_{d}$ the MSE of the channel estimates for the data resource elements. $\sigma_{ICI}^2$ is the average intercarrier interference (ICI) power in the system. User mobility and carrier frequency offset (CFO) are the two major sources of ICI in a wireless system. In our work, the ICI due to CFO is zero since we assume ideal time and frequency synchronization between the transmitter and the receiver. Therefore, user mobility is the only source of ICI in the case of perfect synchronization. The diversity order in a $N_{tx} \times N_{rx}-\text{MIMO}$ system when $N_{tx} \leq N_{rx}$ is given by $\sigma_{ZF}= (N_{rx} - N_{tx} + 1)$ in the absence of antenna correlation \cite{gore_ZF_perf2002}. 
Hence for the SISO and $N_{tx} \times N_{tx}$ MIMO-OFDM ($N_{tx} = N_{rx}$ in our case), $\sigma_{ZF} = 1$. The intercarrier interference power due to user mobility can be upper and lower bounded using \cite{ICI_bounds_Cimini_2001}
\begin{equation}
\label{ICIPower}
\Bigg[ \frac{1}{3} (\pi f_d T_s)^2 -\frac{1}{90} (\pi f_d T_s)^4 \Bigg] \leq \frac{\sigma_{ICI}^2}{\sigma_d^2} \leq \Bigg[ \frac{1}{3} (\pi f_d T_s)^2 \Bigg].
\end{equation}

Note that the expression forming the lower bound in (\ref{ICIPower}) will have to be used in equation (\ref{optimization_problem}) because we are optimizing the upper bound of the achievable rate. 

\begin{figure}[t]
\centering
\includegraphics[width=3.2in]{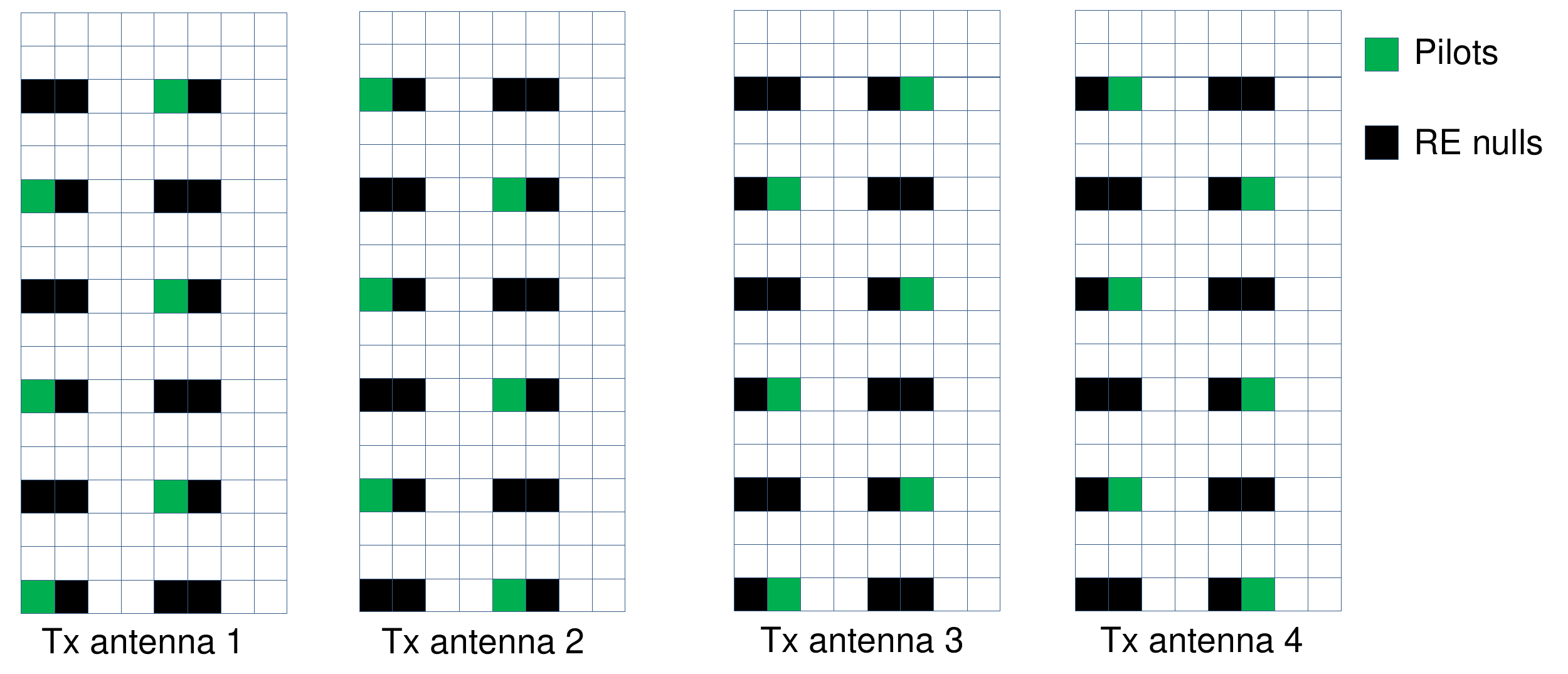}
\caption{Diamond-shaped OFDM pilot arrangement for $4 \times {N_{rx}}$ MIMO-OFDM.}
\label{Fig10_MIMO_pilot_pattern}
\end{figure}

The channel estimation MSE $\delta_d $ will not be known to the receiver, but rather, needs to be estimated. The spectral utilization function depends on the number of data resource elements $N_d$, which are limited only by the  number of pilot REs $N_p$, which in turn depend on the pilot spacing $\Delta_p t$ and $\Delta_p f$. The instantaneous spectrum utilization function is given as 
\begin{equation}
\label{spec_util_func}
S(\Delta_p f, \Delta_p t) = \frac{N_d}{N_d + N_p}.
\end{equation}
For $N$ subcarriers per OFDM symbol with the diamond-shaped pilot arrangement, there will be $N_{f1}$ and $N_{f2}$ pilots in alternate pilot-bearing OFDM symbols. In this paper, we define ``resource block'' to be a collection of contiguous resource elements such that the pilot density across all such blocks is uniform. The number of pilots per resource block of $2N\Delta_p t$ resource elements is $N_p = N_{f1} + N_{f2}$ where $N_{f1} = \lceil N/\Delta_p f \rceil$ and
\begin{align}
 N_{f2} = 
 \begin{cases}
 \lceil N/\Delta_p f \rceil & \text{if } N \Mod{\Delta_p f} > \Delta_p f/2 \\
 \lfloor N/\Delta_p f \rfloor & \text{if } N \Mod{\Delta_p f} \leq \Delta_p f/2,
\end{cases}
\end{align}

\noindent where $N_d$ can be obtained by seeing that in a resource block of $N = (2 N \Delta_p t)$ resource elements, $N_p$ of them are occupied by pilots. If it is a MIMO system, then RE nulls would be necessary to transmit pilot from other antennas, as shown in Fig. \ref{Fig10_MIMO_pilot_pattern}. Therefore, for a $N_{tx} \times N_{rx}-\text{MIMO}$ system, $N_d = (2  N \Delta_p t - N_{tx} N_p )$ and 

\begin{equation}
\label{spec_util_func_final}
S(\Delta_p f, \Delta_p t) = \frac{2 N \Delta_p t - N_{tx} N_p}{2 N \Delta_p t}.
\end{equation}

When we have a average power per RE of $\bar{P}_t$, $N_d \sigma_d^2 + N_p \sigma_p^2 = 2N \bar{P}_{t} \Delta_p t $. For a fixed $\rho$, the data and pilot powers can be obtained as 
\begin{align}
\sigma_d^2 & = \frac{2N \bar{P}_{t} \Delta_p t}{N_p/\rho + N_d} \\
\sigma_p^2 & = \frac{2N \bar{P}_{t} \Delta_p t }{N_p + \rho N_d}.
\label{pilot_data_power_expr}
\end{align}
With this formulation, we still need to estimate some of the terms necessary to calculate (\ref{posteq_SINR}). These quantities are 
\begin{enumerate}
\item Channel estimation  mean square error (MSE) $\delta_d$.
\item $f_d$ in order to estimate the lower bound of $\sigma_{ICI}^2$ and $R_t (\Delta t)$.
\item Channel correlation functions $R_f (\Delta f)$ and $R_t (\Delta t)$ in order to estimate the MSE $\delta_d$.
\item Noise power $\sigma_w^2$.
\end{enumerate}
The estimation of these parameters are outlined in the next two sections.

\section{Channel Estimation MSE}\label{ChEst_MSE}
In this section, we derive closed form expressions for the channel estimation MSE for `diamond shaped' OFDM pilot configurations in doubly selective wireless channels. These expressions help in estimating the MSE due to imperfect channel estimation for a fixed pilot configuration, which is a factor that contributes significantly to the capacity of the OFDM system. 
\subsection{Channel Model}
We model the frequency selectivity of the wireless channel using a tapped-delay line model and temporal variations using the Jake's model \cite{Pil_sym_aid_chanest_Li_2000}. 
We consider a wireless channel under the `Wide Sense Stationary Uncorrelated Scattering' (WSSUS) approximation where the channel correlation $R_H (\Delta f, \Delta t)$ can be simplified as $R_H(\Delta t,\Delta f)=\sigma_H^2 R_t(\Delta t) R_f(\Delta f)$ \cite{Pil_sym_aid_chanest_Li_2000}. $R_t(\Delta t)$ is the channel temporal correlation function and $R_f(\Delta f)$ the spectral correlation function. For simplicity, we assume a channel with $\sigma_H^2 = 1$. The temporal correlation is given by Jake's model using $R_t (\Delta t) = J_0 (2 \pi f_d \Delta t)$ where $J_0(.)$ is the Bessel function of the first kind of zeroth order and the maximum doppler frequency $f_d = vf_c/c$ with $v$ being the relative speed between the receiver and the transmitter, $f_c$ the carrier frequency and $c$ the speed of light.

\begin{figure}[t]
\centering
\includegraphics[width=3.0in]{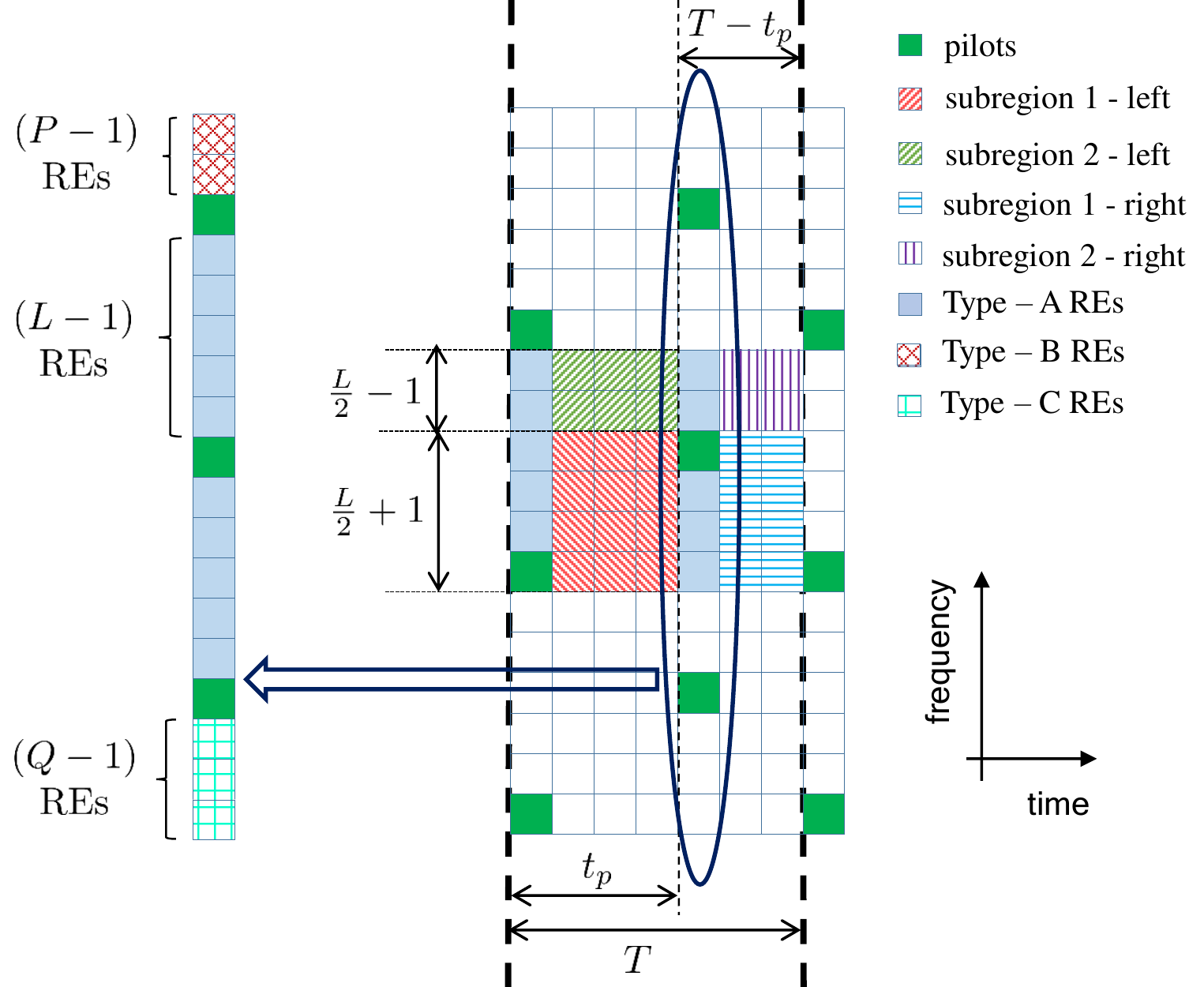}
\caption{Diamond-shaped OFDM pilot arrangement for channel estimation MSE analysis.}
\label{Fig2_MSE_time_interp}
\end{figure}

\subsection{Analysis Region}
To simplify the performance analysis, we divide the OFDM block into four distinct types of resource elements:
\begin{enumerate}
\item Pilots: Their channel estimates are obtained using Least Squares (LS) channel estimation, as shown in equation (\ref{least_squares}).
\item Type A: Resource Elements that lie between 2 pilot subcarriers. Their channel estimates are obtained by interpolation of channel estimates in frequency, between these two pilot subcarriers, as shown in equation (\ref{interp_reg1_l}) with $t = 0$.
\item Type B and C: REs that lie after the last pilot subcarrier (Type B), or before the first pilot subcarrier (Type C). Their channel estimates are obtained by extrapolation of channel estimates in frequency, using the ultimate and penultimate pilots (Type B) and the first and second pilots (Type C). Since they are very few in number, they can been ignored in this analysis. The MSE analysis for these REs are similar to what is presented for Type-A subcarriers.
\item Subregions 1 and 2 : Resource elements that lie between two pilot-bearing OFDM symbols. Their channel estimates are obtained by linear interpolation in frequency and time, as given by equations (\ref{interp_reg1_l})-(\ref{interp_reg2_l}), for $t \neq 0$.
\end{enumerate}
\subsection{Channel Estimation}
Fig. \ref{Fig2_MSE_time_interp} shows the time-frequency resource grid, consisting of resource elements (REs), where the pilot symbols are located on the OFDM symbols at time $(n_1 t_p + n_2 T)$ seconds such that $n_2 \in \mathbb{Z} \text{ and } n_1 \in \{0, 1\} $. The pilot spacing is $\Delta_p f = L$ subcarriers on the frequency axis on the same OFDM symbol, with a relative cyclic frequency shift of $L/2$ between two consecutive pilot-bearing OFDM symbols. 

Let $\mathcal{P}_{ref}$ be set of pilot locations in an OFDM symbol. Let its elements form an ordered pair given by $(l,n) \in \mathcal{P}_{ref}$, where $l$ is the subcarrier index of the pilot at time $n$. Let set $\mathcal{S}$ contain all possible time-frequency locations in the OFDM block. For the pilot at the location $(l,n)$, the LS channel estimate $\hat{H}_l [n]$ will be
\begin{align}
\label{least_squares}
\hat{H}_l[n] & =  \frac{Y_l[n]}{P_l[n]} = H_l[n] + \frac{w_l[n]}{P_l[n]},
\end{align}
where the overall noise $w_l [n]$ can be expressed as a sum of AWGN and ICI components $w_l [n] = w_l^{(AWGN)} [n] + w_l^{(ICI)} [n]$. We consider that $w_l^{(AWGN)} [n] \sim \mathcal{CN}(0, \sigma_w^2)$, $\mathbb{E} [w_l^{(ICI)} [n]] = 0$ and $\mathrm{Var}[w_l^{(ICI)}[n]]=\sigma_{ICI}^2$, where $\sigma_w^2$ is the average noise power and $\sigma_{ICI}^2$ the average ICI power. The channel estimates of the data resource element at the location $(k, n)$ in the left part of subregion 1 is given by interpolation along the time and frequency axes
\begin{align}
\label{interp_reg1_l}
\hat{H}_{k} [n+t]& = \eta \Big[ \Big( \frac{1}{2}-\zeta \Big) \hat{H}_{\frac{-L}{2}}[n+t_p]+ \Big( \frac{1}{2}+\zeta \Big) \hat{H}_{\frac{L}{2}}[n+t_p] \Big] \nonumber \\
&+(1-\eta) \big[ (1-\zeta) \hat{H}_{0}[n] + \zeta \hat{H}_{L}[n] \big],
\end{align}
\noindent where $\eta \triangleq t/t_p$ and $\zeta \triangleq k/L $, for $0 \leq t <t_p$ and $0 \leq k \leq L/2$. Similarly for $0 \leq t < t_p$ and $L/2 \leq k < L$ channel estimates are given by 
\begin{align}
\label{interp_reg2_l}
\hat{H}_{k} [n+t]& = \eta \Big[ \Big( \frac{3}{2}-\zeta \Big) \hat{H}_{\frac{L}{2}}[n+t_p]+ \Big(\zeta - \frac{1}{2} \Big) \hat{H}_{\frac{3L}{2}}[n+t_p] \Big] \nonumber \\
&+(1-\eta) \big[ (1-\zeta) \hat{H}_{0}[n] + \zeta \hat{H}_{L}[n] \big],
\end{align}
\subsection{MSE Analysis}\label{AppendixA}
Fig. \ref{Fig2_MSE_time_interp} shows the analysis region (marked by the colored regions) consisting of subregions 1 and 2, Type-A REs and pilots. Because of the periodic distribution of pilots, the performance in this region will statistically be the same as that of the entire OFDM block. Hence, we derive expressions for the average channel estimation MSE of the REs in this analysis region. 

The average MSE can computed as $\delta_{avg} = \frac{1}{L \cdot T} \mathop{\sum \sum}_{(k,n) \in \mathcal{A}} \mathbb{E}[ \lvert H_k [n] - \hat{H}_k [n] \rvert ^2]$ , where $\mathcal{A}$ denotes the set containing locations of the REs in the analysis region. This can be expressed as a weighted mean of the MSE of the different RE types. 

\subsubsection{MSE of Pilots}
For pilots, the channel estimates are given by (\ref{least_squares}). We consider the ICI term to be uncorrelated with the AWGN term and hence we have $\mathrm{Var}[w_l [n]/P_l[n]] = \frac{\sigma_w^2 +\sigma_{ICI}^2}{\sigma_p^2}$ for $(l,n) \in \mathcal{P}_{ref}$, where $\sigma_p^2$ is the pilot signal power. Furthermore, we consider that the ICI term is uncorrelated with the channel coefficient $H_l [n]$, so that $ \mathbb{E} [w_l [n] H^*_l[n]] = 0 \text{ for }(l,n) \in \mathcal{S}$. The MSE of the pilot channel estimates can be given as 
\begin{align}
\label{MSE_pilot}
\delta_p & = \frac{1}{\lvert \mathcal{P} \rvert} \sum_{(l,n) \in \mathcal{P}} \mathbb{E}[ \lvert H_l [n] - \hat{H}_l [n] \rvert ^2] = \frac{\sigma_w^2 + \sigma_{ICI}^2 }{\sigma_p^2}.
\end{align}
\subsubsection{MSE of Type-A REs}
The Mean Square Error of the channel estimates for Type A REs, denoted by $\delta_{f,A}$, is derived in \cite{Daesik2005}. 
Using our notation it can be represented as
\begin{align}
\label{typeA}
\delta_{f,A} & = \Big (\frac{5L-1}{3L} \Big ) R_f(0) +   \Big (\frac{2L-1}{3L} \Big ) \Big( \frac{\sigma_w^2 + \sigma_{ICI}^2}{\sigma_p^2} \Big) \nonumber \\
& + \Big (\frac{L+1}{3L} \Big ) \Re(R_f(L)) + \gamma,
\end{align}
where $\gamma = -\frac{2}{L-1}\sum_{i=1}^{L-1} \big[ \big (\frac{L-i}{L} \big ) \Re(R_f(i)) + \frac{i}{L} \Re(R_f(i-L)) \big]$ represents the residual terms.
\subsubsection{Left Part of Subregion 1 $(0 \leq k \leq L/2, 1 \leq t < t_p )$}
For this subregion, the MSE expression for linear interpolation using Least Squares $\delta_{1,l}$, is
\begin{align}
\label{MSE_t_reg1_l}
\delta_{1,l}=C_1 \sum_{k=0}^{L/2} \sum_{t=1}^{t_p-1} \mathbb{E} \{ \lvert \hat{H}_{k} [n+t] 
- H_{k}[n+t] \rvert ^2 \},
\end{align}
 where $C_1 \triangleq \frac{1}{(L/2+1)(t_p-1)}$. After expanding the terms and simplifying, we get
\begin{align}
\label{reg1_MSE_left_full}
\delta_{1,l}&= (1+\lambda \omega) R_f(0) R_t(0) +  \lambda (2-\omega) R_t(0) \Re(R_f(L)) \nonumber \\
& +(1 - 2\lambda) R_t(t_p) \Re \Big[ \omega' R_f \Big( \frac{L}{2} \Big) + (1-\omega') R_f \Big(\frac{3L}{2} \Big) \Big]  \nonumber \\
& +\lambda \omega \Big( \frac{\sigma_w^2 + \sigma_{ICI}^2}{\sigma_p^2} \Big) - \varepsilon_{1,l},
\end{align}
where $\lambda \triangleq \frac{2t_p - 1}{6t_p} ; \omega \triangleq \frac{4L+1}{3L} ;\omega' \triangleq \frac{23L+2}{24L}$ and the cross terms $\varepsilon_{1,l}$ is given by
\begin{align}
\label{crossterms_t_reg1_l}
\varepsilon_{1,l} & = 2 C_1 \sum_{k=0}^{L/2} \sum_{t=1}^{t_p-1} \Big \{(1 - \eta) R_t(t) \Re\Big[ (1-\zeta) R_f(k)  \nonumber \\
& +\zeta R_f(L-k) \Big] + \eta R_t(t-t_p) \Re \Big[ \Big(\frac{1}{2}-\zeta \Big) \cdot \nonumber \\
& R_f \Big(\frac{L}{2}+k \Big) + \Big(\frac{1}{2} + \zeta \Big) R_f \Big(k-\frac{L}{2} \Big) \Big] \Big \}.
\end{align}
\subsubsection{Left Part of Subregion 2}
The MSE for the left part of subregion 2, $\delta_{2,l}$, can be evaluated similarly as shown in equations (\ref{MSE_t_reg1_l})-(\ref{crossterms_t_reg1_l}). 
\subsubsection{Right Parts of Subregion 1 and 2}
For the right part of subregions 1 and 2, the MSEs $\delta_{1,r}$ and $\delta_{2,r}$ can be obtained by taking $t \rightarrow -t$ and $t_p \rightarrow (T - t_p)$ appropriately. $R_t (\Delta t) = R_t (-\Delta t)$ since $J_0 (.)$ is an even function. Therefore, the MSE expressions will take a similar form as (\ref{MSE_t_reg1_l})-(\ref{crossterms_t_reg1_l}). The expressions for $\delta_{1,r} \text{ and } \delta_{2,r}$ have been omitted owing to similarity in the approach and functional form. 

\subsubsection{Average MSE}
The average MSE $\delta_{avg}$ will be the weighted mean of the MSEs of the different RE types in the analysis region. 
\begin{align}
\label{MSE_avg}
\delta_{avg} = & \frac{1}{L \cdot T} \Bigg[ \frac{\delta_{1,l}}{C_1} + \frac{\delta_{2,l}}{C_2} + \frac{\delta_{1,r}}{C_3} + \frac{\delta_{2,r}}{C_4} + C_5 \delta_{f,A} + 2\delta_p  \Bigg],
\end{align}
where $C_2 = \frac{1}{(L/2 - 1)(t_p - 1)}, C_3 = \frac{1}{(L/2 + 1)(T - t_p - 1)}, C_4 = \frac{1}{(L/2 - 1)(T - t_p - 1)}  \text{ and } C_5 = 2(L-1)$. 

For symmetric pilot spacing $T = 2t_p, \delta_{1,l} = \delta_{1,r}, C_1 = C_3 \text{ and } C_2 = C_4$ in (\ref{MSE_avg}). Therefore the MSE of the data REs $\delta_d$ will be given by
\begin{align}
\label{MSE_data}
\delta_{d} = \frac{2}{(L \cdot T - 2)} \Bigg[ \frac{\delta_{1,l}}{C_1} +  \frac{\delta_{2,l}}{C_2} + (L - 1)\delta_{f,A} \Bigg].
\end{align}

\section{Optimal Pilot Spacing and Power}\label{opt_pil_spac_pow}
\subsection{Estimation of Parameters}
Noise power can be estimated using the methods proposed in \cite{Cui_PDP_NoiseVarEst_2006, Xu_Zhu_SNREst_OFDM_2005}. To estimate the channel statistics $\hat{R}_t (\Delta t)$ and $\hat{R}_f (\Delta f)$ in a nonstationary wireless channel, temporal averaging can be performed assuming local stationarity of the channel for the averaging duration \cite{Dietrich_chan_sec_stat_estim_2005}. For a $N \times T_{ofdm}$ channel matrix $\hat{\mathbf{H}}$ with $N$ rows corresponding to frequency subcarriers, and $T_{ofdm}$ columns corresponding to OFDM symbols, the fading statistics can be estimated using
\begin{align}
\label{estimate_statistics}
\hat{R}_t (-i) &= \frac{1}{T_{ofdm} - \lvert i \rvert} \sum_{t = 1}^{T_{ofdm} - \lvert i \rvert} \Big \{\text{diag}_i \Big[ \hat{\mathbf{H}}^H \hat{\mathbf{H}}  \Big] \Big \}_t  \nonumber \\
\hat{R}_f (-j) &= \frac{1}{N - \lvert j \rvert} \sum_{f = 1}^{N - \lvert j \rvert} \Big \{ \text{diag}_j  \Big[ \hat{\mathbf{H}} \hat{\mathbf{H}}^H  \Big] \Big \}_f ,
\end{align}
where $\text{diag}_i [\mathbf{X}]$ is the vectorized $i^{th}$ diagonal of matrix $\mathbf{X}$ and $\Big \{\text{diag}_i [\mathbf{X}] \Big \}_k$ its $k^{th}$ element. Because $\hat{\mathbf{H}}^H \hat{\mathbf{H}}$ and $\hat{\mathbf{H}} \hat{\mathbf{H}}^H$ are Hermitian-symmetric matrices, the other elements can be found using $\hat{R}_t (-i) = \hat{R}^*_t (i)$ and $\hat{R}_f (-j) = \hat{R}^*_f (j)$. Using equation (\ref{estimate_statistics}), we form the channel correlation vectors 
\begin{align}
\mathbf{\hat{R}_f} & =\Big[\hat{R}_f ({\tfrac{-N_{\Delta f}}{2}}) \cdot \cdot \hat{R}_f(-1)\ \hat{R}_f(0)\ \hat{R}_f(1) \cdot \cdot \hat{R}_f ({\tfrac{N_{\Delta f} - 2}{2}}) \Big], \nonumber \\
\mathbf{\hat{R}_t} & =\Big[\hat{R}_t ({\tfrac{-N_{\Delta t}}{2}} ) \cdots \hat{R}_t(-1)\ \hat{R}_t(0)\ \hat{R}_t(1) \cdots \hat{R}_t ({\tfrac{N_{\Delta t} - 2}{2}} ) \Big].
\end{align}
Without loss of generality, we assume that the vector lengths $N_{\Delta f} \text{ and } N_{\Delta t}$ are positive even integers. In practical scenarios where the channel statistics are estimated over a finite duration, the accuracy will be poor. This occurs due to (a) interpolation error, and (b) addition of noise. In the worst case, the estimated channel statistics can violate the properties of the autocorrelation function $\vert \hat{R}_t (\Delta t) \rvert \leq \hat{R}_t (0)\ \forall\ \Delta t \neq 0 $. This can happen especially in high noise, low mobility and/or flat fading scenarios. Using these estimated channel statistics directly can result in inconsistent and, sometimes absurd values for the MSE. Therefore, we propose a codebook-based approach to increase the robustness of the feedback. The codebook contains the power delay profile (PDP) and maximum Doppler frequency values of typical channels that the radio expects to encounter. A cognitive radio, for example, can update the codebook over time as it learns more about its channel environment. The receiver calculates the channel statistics using equation (\ref{estimate_statistics}) for a finite duration and finds the codebook profile that is closest to it in the minimum euclidean distance sense.

\subsection{Channel Statistics Codebook}
Let the codebook be denoted by set $\mathcal{R}_C$ with two disjoint subsets $\mathcal{R}_{C,t} \subseteq \mathcal{R}_C$ and $\mathcal{R}_{C,f} \subseteq \mathcal{R}_C$, where $\lvert \mathcal{R}_{C,f} \rvert = M_f$ and $\lvert \mathcal{R}_{C,t} \rvert = M_t$. $\mathcal{R}_{C,f}$ is the set of channel frequency correlation profiles, with $N_{\Delta f} \times 1$ vector elements $\mathbf{R_{fc,l}} \in \mathcal{R}_{C,f}$ for $1 \leq l \leq M_f$. Likewise, $\mathcal{R}_{C,t}$ is the set of channel temporal correlation profiles, with $N_{\Delta t} \times 1$ vector elements $\mathbf{R_{tc,m}} \in \mathcal{R}_{C,t}$ for $1 \leq m \leq M_t$. Here, we model temporal fading using a classic Doppler spectrum where the $\Delta t^{th}$ element is $[\mathbf{R_{tc,m}}]_{\Delta t} = J_0(2 \pi f_{d,m} \Delta t)$ \cite{jakes1994microwave}. $f_{d,m}$ is the maximum Doppler frequency for the $m^{th}$ temporal correlation profile. Such a definition of the codebook channel profiles is motivated by the WSSUS approximation.

Initially, the profiles that comprise the codebook would correspond to the most common types of channels that the radio would be expected to encounter, based on reported field measurements. For example the channel profiles from ITU-T \cite{ITUIMT2000spec} and the 3GPP channel models \cite{3GPPLTE_TS36_104_tx_and_rx} can be used as initial codebook entries. In the case of a cognitive radio, the codebook can be updated over time, when it learns more about its operating channel environment. The codebook can be designed to match the typical scenarios operation environment of the radios. For example vehicular to vehicular networks would have a large variation in Doppler spreads. On the other hand, UAV-to-UAV systems might have very low root mean square delay spread due to strong line of sight propagation \cite{Malotak_UAVCHanChar_2016, Zeng_UAV_Chall_2016}. We will provide example codebooks in the next section.

\begin{algorithm}[t]
\small
\begin{algorithmic}[1]
 \State \textbf{Input:}
 \Statex \emph{Codebook $\mathcal{R}_C$}
 \Statex \emph{Sets $\mathcal{D}_f, \mathcal{D}_t \text{ and } \mathcal{P}$, that are known to the transmitter.}

\State Estimate $\hat{R}_t$ and $\hat{R}_f$ from equation (\ref{estimate_statistics}) using $\mathbf{\hat{H}}$, computed using the most recent $T_{ofdm}$ OFDM symbols. 

\State Find the frequency and time domain channel profiles from the codebook, $\mathbf{R_{fc, l'}} \in \mathcal{R}_{C,f}$ and $\mathbf{R_{tc, m'}} \in \mathcal{R}_{C,t}$ by solving
\begin{align}
\label{optimize_pilots}
l' = \underset{1 \leq l \leq M_f} {\text{arg min}} \quad
& \| \mathbf{\hat{R}_f} - \mathbf{R_{fc, l}} \| \nonumber \\
m' = \underset{1 \leq m \leq M_t} {\text{arg min}} \quad
& \| \mathbf{\hat{R}_t} - \mathbf{R_{tc, m}} \|. 
\end{align}

\noindent For a $N_{tx} \times N_{rx}$ MIMO-OFDM, there will be $N_{tx} N_{rx}$ channel matrices of dimension $N \times T_{ofdm}$ (one for each transmit-receive antenna pair). If $\mathbf{l'} \text{ and } \mathbf{m'}$ represent the $N_{tx} N_{rx} \times 1$ vectors of codebook indices found using equation (\ref{optimize_pilots}) for each channel matrix, then $l' = \text{mode}(\mathbf{l'}), m' = \text{mode}(\mathbf{m'})$. 

\State Feed back the codebook indices $l' \text{ and } m'$ to the transmitter on the uplink.

\State For $\rho \in \mathcal{P}, \Delta_p f \in \mathcal{D}_f, \Delta_p t \in \mathcal{D}_t$, 
compute channel estimation MSE $\delta_d$ assuming channel statistics $R_{fc, l'}$ and $R_{tc, m'}$ using equation (\ref{MSE_data}).

\State Using the values of $\delta_d$ for each tuple $\{ \rho, \Delta_p f, \Delta_p t \}$, solve equation (\ref{optimize_pilots}) by calculating all the other necessary terms using equations (\ref{posteq_SINR})-(\ref{pilot_data_power_expr}). Let the resulting optimal tuple be $\{ \rho_{o}, (\Delta_p f)_{o}, (\Delta_p t)_{o} \} $.

\State Find the new pilot power and locations using $\{ \rho_{o}, (\Delta_p f)_{o}, (\Delta_p t)_{o} \} $. 

\State For the next $T_{ofdm}$ OFDM symbols received, estimate the channel matrix/matrices $\mathbf{\hat{H}}$. 

\State Go back to step 1.
\end{algorithmic}
\caption{Pilot Adaptation: Receiver Processing}
\label{algo_pilot_adapt}
\end{algorithm}

\begin{algorithm}[t]
\small
\caption{Pilot Adaptation: Transmitter Processing}
\label{algo_pilot_adapt1}
\begin{algorithmic}[1]
\State \textbf{Input: }
 \Statex \emph{Codebook $\mathcal{R}_C$}
 \Statex \emph{Sets $\mathcal{D}_f, \mathcal{D}_t \text{ and } \mathcal{P}$, that are known to the receiver.}
\State Based on the received codebook indices $l' \text{ and } m'$, for $\rho \in \mathcal{P}, \Delta_p f \in \mathcal{D}_f, \Delta_p t \in \mathcal{D}_t$ compute channel estimation MSE $\delta_d$ assuming channel statistics $R_{fc, l'}$ and $R_{tc, m'}$ using equation (\ref{MSE_data}).

\State Using the values of $\delta_d$ for each tuple $\{ \rho, \Delta_p f, \Delta_p t \}$, solve equation (\ref{optimize_pilots}) by calculating all the other necessary terms using equations (\ref{posteq_SINR})-(\ref{pilot_data_power_expr}). Let the resulting optimal tuple be $\{ \rho_{o}, (\Delta_p f)_{o}, (\Delta_p t)_{o} \} $.

\State Find the new pilot power and locations using $\{ \rho_{o}, (\Delta_p f)_{o}, (\Delta_p t)_{o} \} $. 

\State Transmit the next $T_{ofdm}$ OFDM symbols using these new pilot locations and power, on the downlink.

\State Go back to step 1. 
\end{algorithmic}
\end{algorithm}

\subsection{Optimal Pilot Spacing and Power}
\begin{figure}[t]
\centering
\includegraphics[width=3.2in]{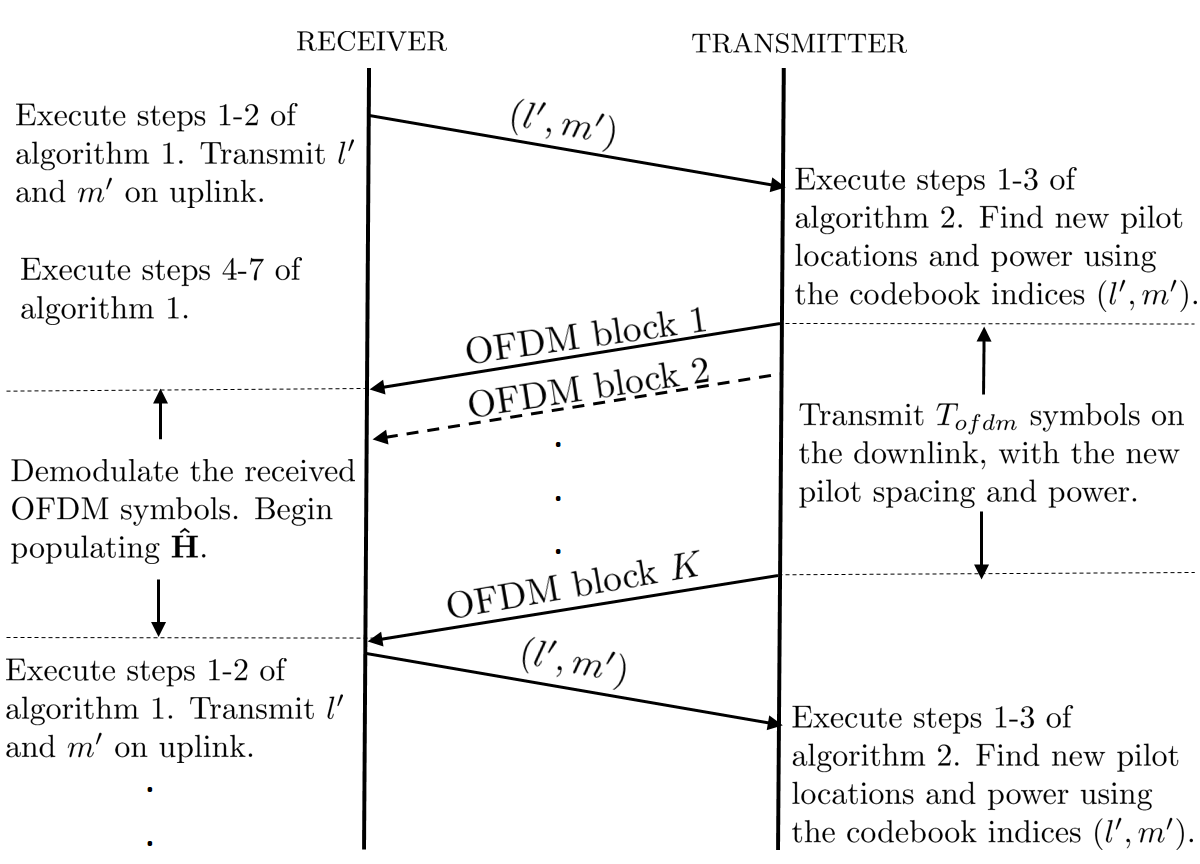}
\caption{Illustration of the typical exchange between the transmitter and the receiver, in our pilot spacing and power adaptation algorithm. $K$ OFDM blocks are equivalent to $T_{ofdm}$ OFDM symbols.}
\label{Fig_Pil_adapt_algo_exchange}
\end{figure}

We assume that the transmitter and receiver both know and share a common $\mathcal{P}, \mathcal{D}_f$ and $\mathcal{D}_t$; the sets that contain allowable values for $\rho, \Delta_p f$ and $\Delta_p t$, respectively. With the range for each parameter predefined based on the constraints in (\ref{optimization_problem}), the algorithms to find the optimal pilot spacing and power can be executed once every $T_{ofdm}$ symbols as shown in algorithm \ref{algo_pilot_adapt} and \ref{algo_pilot_adapt1}. Each algorithm is executed once using the most recent $T_{ofdm}$ symbols. Upon its completion, it uses the subsequent $T_{ofdm}$ symbols for the next cycle of pilot adaptation, and so on. Fig. \ref{Fig_Pil_adapt_algo_exchange} shows the typical sequence in which algorithms \ref{algo_pilot_adapt} and \ref{algo_pilot_adapt1} are executed in the receiver and transmitter respectively.

\subsection{Feedback Requirements and Computational Complexity}
\subsubsection{SISO and MIMO-OFDM systems}
Based on the above algorithm for pilot adaptation, the receiver needs to feed back the indices of the corresponding channel profile from the codebook. There are a total of $M_t M_f$ possible values that can be sent to the transmitter. Therefore, the receiver would need to feed back $b_f = \lceil \log_2 (M_t M_f ) \rceil$ bits. Therefore, $b_f$ bits are exchanged between the transmitter and receiver once in every $(T_{ofdm} \times T_{s})$ seconds leading to a bit-rate of $\frac{b_f}{T_{ofdm} T_s}$ bits per second.

Estimation of channel statistics involve matrix multiplication, which can be accomplished with a complexity of $O(N^2 T_{ofdm})$ for each element of $\mathbf{\hat{R}_f}$, and $O(T_{ofdm}^2 N)$ for $\mathbf{\hat{R}_t}$. Values for $T_{ofdm}$ and $N$ have to be chosen to estimate the channel statistics accurately. Since these operations are similar to those used in an MMSE receiver which relies on accurate estimation of the channel statistics \cite{Arslan_OFDM_chanest_survey_2007}, its implementation does not consume additional computing resources in modern wireless receivers. The other steps involved in algorithm \ref{algo_pilot_adapt} and \ref{algo_pilot_adapt1} are of low complexity and hence do not burden modern wireless radios.

For MIMO-OFDM, the computational complexity to estimate $\mathbf{\hat{R}_t} \text{ and } \mathbf{\hat{R}_f}$ is $O(N_{tx} N_{rx} T_{ofdm}^2 N)$ and $O(N_{tx} N_{rx} N^2 T_{ofdm})$ respectively. The feedback requirements will remain the same as in the case of SISO-OFDM.

\subsubsection{Multi-Band Carrier Aggregation}
In multi-band carrier aggregation, the resource blocks can be allocated to a user across two or more frequency bands. In such a case, pilots will be sent on all $N_{b}$ allocated bands $(f_1, f_2, \cdots f_{N_b})$ and the pilot spacing can be varied on each frequency band based on its channel statistics. In this case, some of the properties of Doppler spread can be exploited to reduce the computation and feedback requirements for the channel profile in the codebook. We assume that the OFDM symbol duration, subcarrier spacing and all other parameters except for the pilot spacing and power, are the same across all frequency bands. Since the Doppler frequency scales linearly with the center frequency $f_c$, only one codebook index specifying the temporal pilot spacing needs to be fed back for any one of the $N_b$ bands. The temporal codebook index $m'$ for the other $(N_b - 1)$ bands can be estimated at the transmitter by back calculations. Even in the case where each frequency band experiences a different root mean square delay spread, the total number of bits needed for feedback will be $b'_f = \lceil \log_2 (M_t M_f + (N_b - 1)\times M_f) \rceil$. Hence with this method, at least $\Big \lceil \log_2 \Big( \frac{N_b M_t M_f}{M_t M_f + (N_b - 1)\times M_f} \Big) \Big \rceil$ bits of feedback can be saved. Similar to the case of single-band OFDM systems, the bit rate requirement to implement adaptive pilot spacing and power is $\frac{b_f'}{T_{ofdm} T_s}$ bits per second.

\subsection{Extension to Other Types of Receivers}
In this work, we have focused on least squares with linear interpolation channel estimation, and ZF equalization. There are more robust methods such as Minimum mean square error (MMSE) and Linear MMSE (LMMSE). The derivation of the mean square error for these estimators is beyond the scope of this paper. We direct interested readers to \cite{vsimko2012optimal} (equations (32)-(36)) for the MSE expression for LMMSE. However, using our framework some simplifications are possible to ease the burden on numerical computation by using the WSSUS approximation for the channel correlation matrices:
\begin{enumerate}
\item The elements of matrices $\mathbf{R_{h_p, h_p}}$, $\mathbf{R_{h_d, h_p}}$ and $\mathbf{R_{h_d, h_d}}$ (in equation (36) of \cite{vsimko2012optimal}) take the form $R_t(\Delta t) \cdot R_f(\Delta f)$ when simplified using the WSSUS model.
\item We can use the codebook to populate the correlation matrices, since channel statistics vary fairly slowly. After finding $l'$ and $m'$ in equation (\ref{optimize_pilots}), we can use the codebook entries to rapidly compute the channel autocorrelation and crosscorrelation matrices and hence, the MSE for each pilot configuration.
\item $\sigma_{ICI}^2$ can be directly obtained using our codebook $\mathcal{R}_{C,t}$, using equation (\ref{ICIPower}).
\end{enumerate}

It is important to note that our scheme is general and can be used with any OFDM channel estimator and equalizer when the channel estimation MSE, ICI power and the diversity order per stream $\sigma$ ($\sigma_{ZF}$ in this paper) can be estimated with a reasonable accuracy at the receiver. 

\begin{figure}[t]
\centering
\includegraphics[width=3.0in]{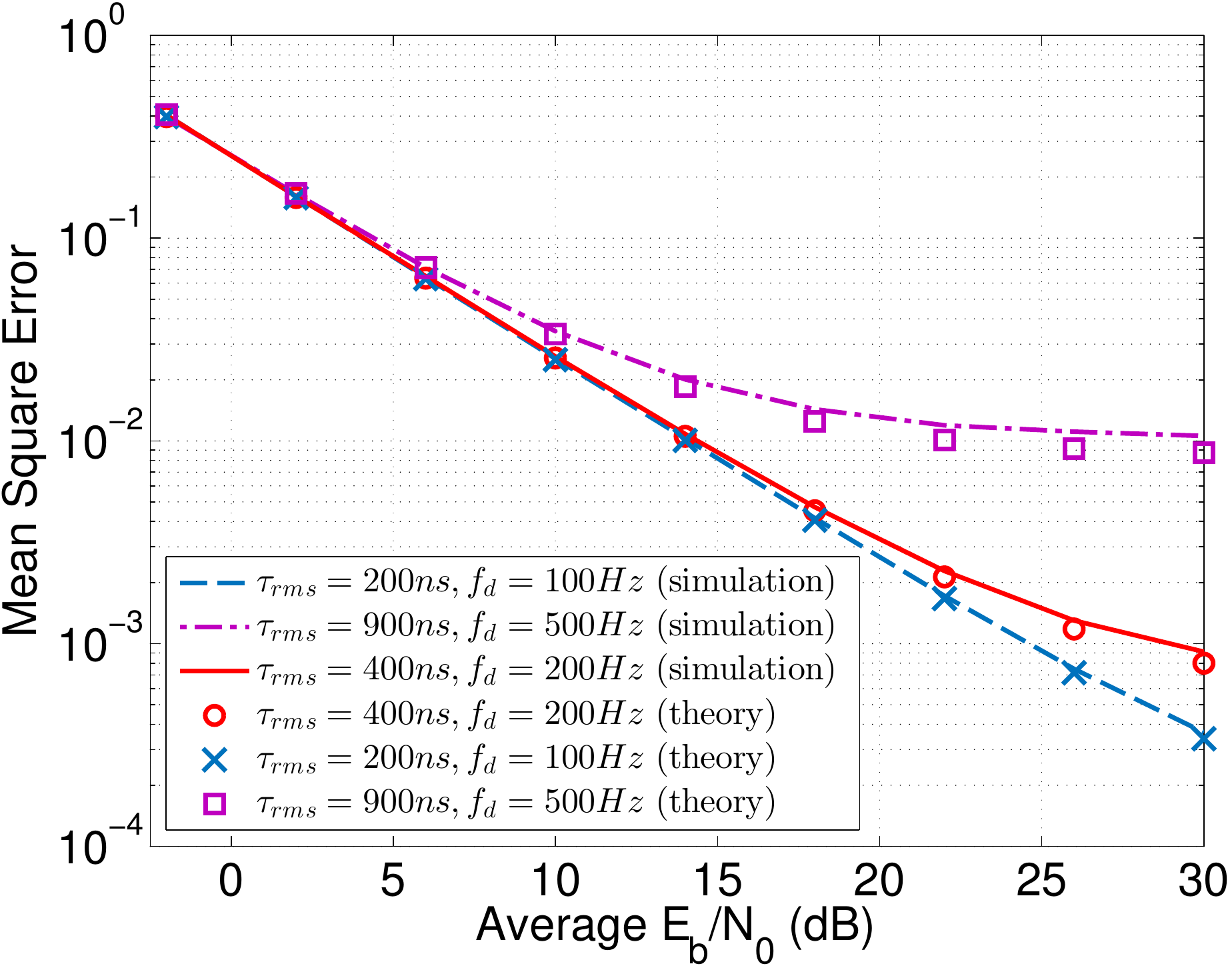}
\caption{Comparison of theoretical and simulated channel estimation MSE values for different channel conditions for SISO-OFDM.}
\label{Fig3_MSE_curves}
\end{figure}

\begin{table}[t]
\renewcommand{\arraystretch}{1.0}
\caption{Simulation Parameters}
\label{Tab2_OFDM_params}
\centering
\begin{tabular}{|l|l|}
\hline
\textbf{Parameter} & \textbf{Value} \\
\hline
Antenna Configuration & SISO and $4\times 4$ MIMO \\
\hline
FFT-length & 128\\
\hline
Number of OFDM subcarriers & 72 \\
\hline
Number of Guard Subcarriers & 28 on each band edge\\
\hline
Center Frequency $f_c$ & $700\ \text{MHz}$ and $2.0\ \text{GHz}$ \\
\hline
Subcarrier Spacing $f_{sub}$ & $15\ \text{kHz}$\\
\hline
OFDM symbol duration $T_s$ & $71.875\ \mu s$\\
\hline
Cyclic Prefix Duration & $5.21\ \mu s$\\
\hline
Base pilot spacing in time $t_p$ & 4 OFDM symbols $(0.2875\ \text{ms})$\\
\hline
Base pilot spacing in frequency $L$ & 6 subcarriers $(90\ \text{kHz})$\\
\hline
Channel parameters & Doubly selective: Jakes Doppler \\
 & spectrum with multipath fading.\\
\hline
Vehicular speed & 0-500 km/h \\
\hline
rms delay spread & 0-300 ns (air-to-ground) \\
& 0-1000 ns (V2V) \\
\hline
Channel Estimation & Least Squares (pilots) \\
& 2D-Linear Interpolation (data REs)\\
\hline
Equalization & Zero Forcing (ZF) \\
\hline
\end{tabular}
\end{table}

\section{Numerical Results}\label{num_results}
We present the numerical results in this section. We simulated an OFDM system in a doubly selective fading channel. Jakes Doppler Spectrum models the mobility effects in the channel, with Rayleigh fading due to multipath modeled using a tapped delay-line model. The parameter $\tau_{rms}$ controls the frequency selectivity of the channel and $f_d$ the maximum Doppler frequency. Table \ref{Tab2_OFDM_params} summarizes the parameters of the OFDM waveform as illustrated in Fig. \ref{Fig1_pilot_adapt_illustrate} for SISO-OFDM and Fig. \ref{Fig10_MIMO_pilot_pattern} for MIMO-OFDM. We also simulated CA-OFDM systems with two component subcarriers at $f_1  = 700 \text{ MHz and } f_2 = 2 \text{ GHz}$. 
\subsection{Channel Estimation MSE Performance}
Fig. \ref{Fig3_MSE_curves} shows the comparison between the theoretical and simulated channel estimation performance for doubly selective channels of different characteristics. We have computed the ICI power using the lower bound in equation (\ref{ICIPower}). We see that the curves match well, validating the derived MSE expressions in equations (\ref{least_squares})-(\ref{MSE_avg}) in section \ref{ChEst_MSE}. We see that there is a slight mismatch at higher values of $E_b/N_0$ as $f_d$ increases, because the ICI power that we have considered in our theoretical expression is approximate. The contribution of ICI becomes prominent at higher $E_b/N_0 \text{ and } f_d$. The deviation is negligible in relatively low frequency selective and mobility conditions. Even in high mobility conditions ($f_d = 500 \text{ Hz}$), the theoretical expressions form a tight lower bound to the channel estimation MSE. Hence our derived MSE expressions can be used to maximize the upper bound of the achievable rate in algorithms \ref{algo_pilot_adapt} and \ref{algo_pilot_adapt1}. 


\begin{figure}[t!]
    \centering
    \begin{subfigure}[t]{0.24\textwidth}
    \label{7a}
        \raggedleft
        \includegraphics[width=1.7in]{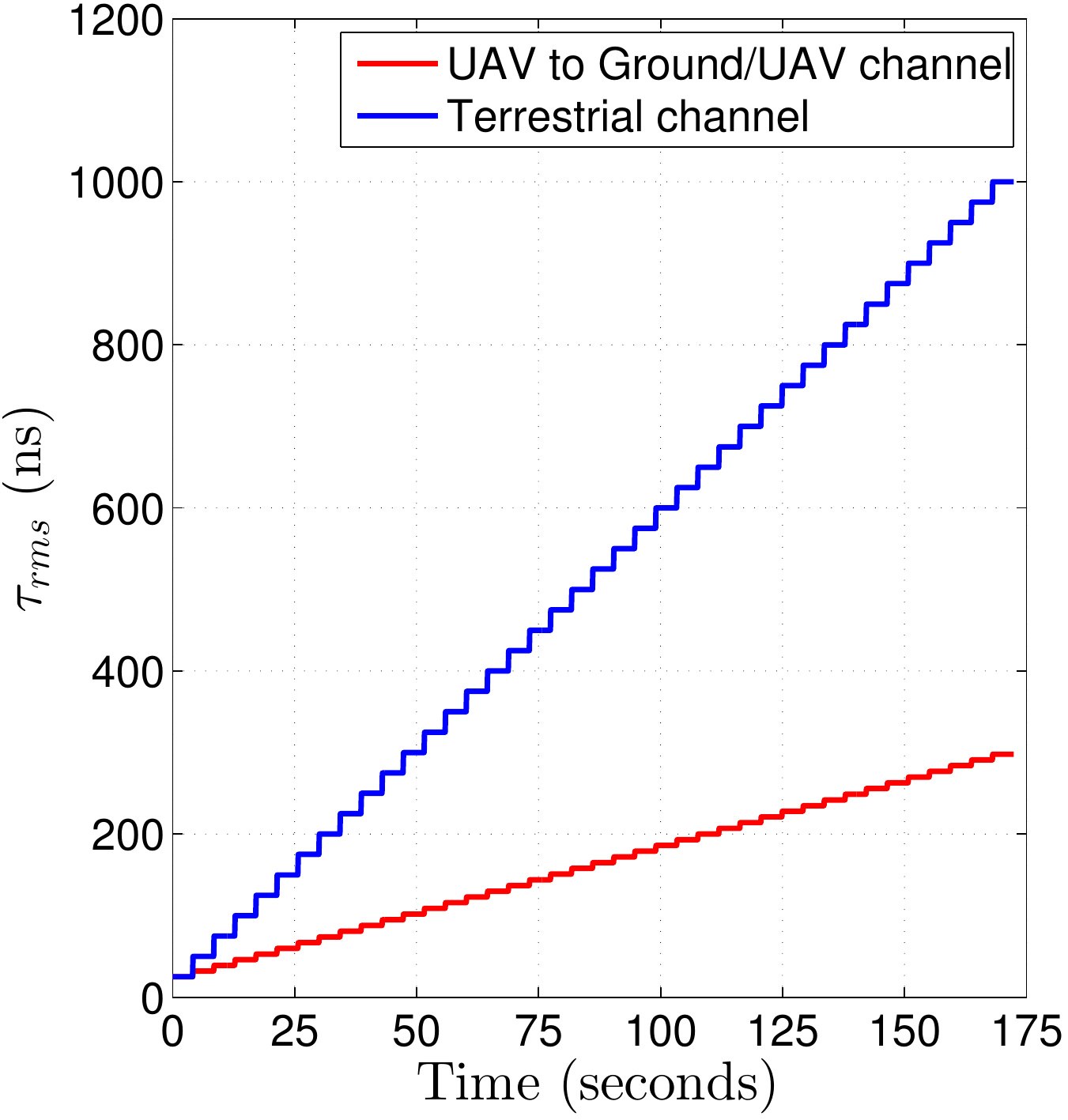}
        \caption{}
    \end{subfigure}%
    ~ 
    \begin{subfigure}[t]{0.24\textwidth}
    \label{7b}
        \centering
        \includegraphics[width=1.7in]{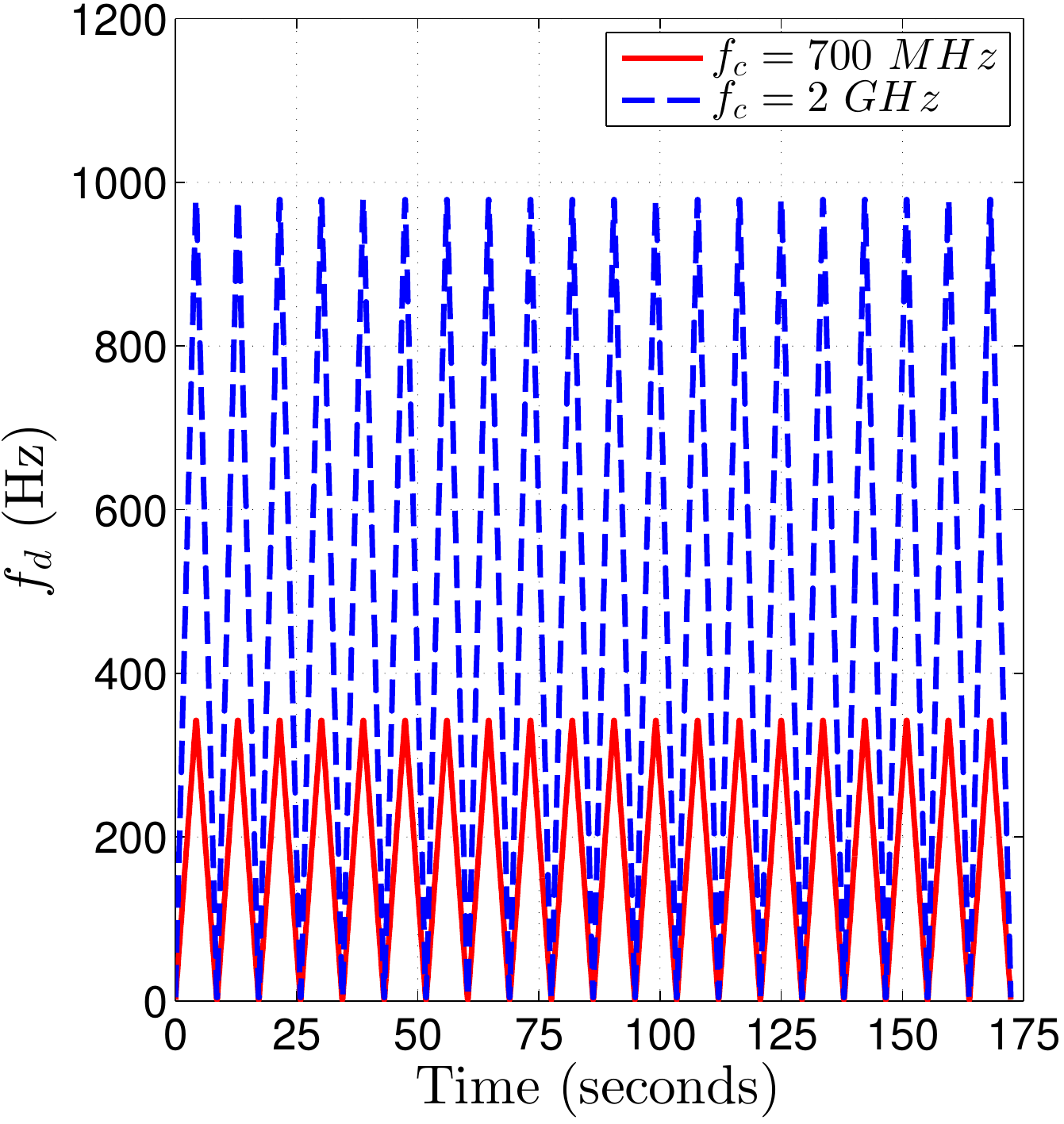}
        \caption{}
    \end{subfigure}
    \caption{Variation of the (a) root mean square delay spread and (b) maximum Doppler frequency, of the simulated doubly selective nonstationary wireless channel scenarios.}
    \label{Fig7_channel_stats_varn_all}
\end{figure}

\subsection{Pilot Adaptation in Doubly Selective Nonstationary Wireless Channels} \label{nonCAResults}
We simulate a doubly selective nonstationary wireless channel with slowly varying second order statistics, as illustrated in Fig. \ref{Fig7_channel_stats_varn_all}. We consider the following channel scenarios: (a) UAV to ground wireless channel and (b) terrestrial wireless channel. UAV-to-UAV or UAV-to-ground wireless channels are typically characterized by a low $\tau_{rms}$ when compared to terrestrial channels \cite{Malotak_UAVCHanChar_2016, Zeng_UAV_Chall_2016}. 

We model $\tau_{rms}$ to be the same for both frequency bands at any given time. On the other hand, the maximum Doppler frequency $f_d$ is directly proportional to $f_c$, as shown in Fig. \ref{Fig7_channel_stats_varn_all}. For adaptive pilot configurations, the parameter ranges are $\rho \in \mathcal{P}=\{-9 \text{ dB}, -8\text{ dB}, \cdots, 0\text{ dB} \}, \Delta_p t \in \mathcal{D}_t=\{2, 3, \cdots, 10 \} \text{ and } \Delta_p f \in \mathcal{D}_f = \{2, 4, \cdots, 12\}$. The channel statistics are estimated once every $T_{ofdm} = 1500$ OFDM symbols over $N=72$ subcarriers. 

On the other hand, channel estimation needs to be performed before any data symbol can be decoded by the receiver. Since we are using linear interpolation between two pilot-carrying OFDM symbols, this operation needs to be performed once every $\Delta_p t \times T_{s}$ seconds in our work, where $\Delta_p t $ is the pilot spacing in time, and $T_{s}$ the OFDM symbol duration. Since $ 2 \leq \Delta_p t \leq 10$ and $T_{s} = 71.875 \mu s$, we perform channel estimation once every $144 \text{ to } 719 \mu s$, depending on the value of $\Delta_p t$. Popular wireless standards such as LTE and WiFi perform channel estimation on a similar timescale.

For the channel scenarios shown in Fig. \ref{Fig7_channel_stats_varn_all} we compare the throughput performance of adaptive pilot configuration against the following fixed pilot configurations: (a) $\Delta_p t = \Delta_p f = 6, \rho = -3 \text{ dB}$, (b) $\Delta_p t = \Delta_p f = 8, \rho = -3 \text{ dB}$ and (c) pilot configuration of Normal-Cyclic Prefix (CP) LTE \cite{sesia2011lte} with $\rho = -3 \text{ dB}$. We consider both SISO-OFDM and $4 \times 4$ MIMO-OFDM with full spatial multiplexing. 

The designed codebook to implement adaptive pilot configurations using algorithms \ref{algo_pilot_adapt} and \ref{algo_pilot_adapt1}, is shown in Table \ref{Tab6_codebook_profiles}. The codebook vectors lengths in $\mathcal{R}_{C,f}$ and $\mathcal{R}_{C,t}$ are $N_{\Delta f}=62 \text{ and } N_{\Delta t}=40$ respectively. The channel profiles correspond to standard 3GPP and ITU-T channel models \cite{ITUIMT2000spec, 3GPPLTE_TS36_104_tx_and_rx}, and additional codebook entries ensure that the entire range of $\tau_{rms}$ and $f_d$ is efficiently handled by the adaptive OFDM waveform. 

\begin{table}[!t]
\renewcommand{\arraystretch}{1}
\caption{Codebook of Channel Profiles, $\mathcal{R}_C$}
\label{Tab6_codebook_profiles}
\centering
\subcaption*{A: Channel profiles for Doppler Frequency $(\mathcal{R}_{C,t})$}
\begin{tabular}{|c|l|l|}
\hline
Codebook & Mobility Type/Velocity$^\dagger$ & $f_d$ (Hz)\\
Index $(m)$ & &  \\
\hline
1 & Pedestrian (3km/hr) & 5.6\\
\hline
2 & Urban Vehicular (32km/hr) & 60\\
\hline
3 & Highway Vehicular (120km/hr) & 222.22\\
\hline
4 & High Speed Train/UAV low (300km/hr) & 555.56\\
\hline
5 & High Speed Train/UAV medium (400km/hr) & 750\\ 
\hline
6 & High Speed Train/UAV high (500km/hr) & 925\\
\hline
\end{tabular}
\bigskip
\subcaption*{B: Channel profiles for Frequency Selectivity $(\mathcal{R}_{C,f})$}
\begin{tabular}{|c|c|c|c|}
\hline
Codebook & Normalized PDP & Delay & $\tau_{rms}$ \\
Index $(l)$ & & taps* & (ns) \\
\hline
1 & [0.9310, 0.3425, 0.126] & [0,1,2] & 221.5\\
\hline
2 & [0.8882, 0.3152, 0.2809,  & [0,1,2,3,5] & 476.4\\
  & 0.158, 0.0888] & & \\
\hline 
3 & [0.778, 0.4426, 0.3097, & [0,1,2,4,7] & 791.2\\
& 0.3169, 0.0497]  & & \\
\hline
4 & [0.5795, 0.4745, 0.3885,  & [0,1,2,3,4, & 1440\\
 & 0.318, 0.2604, 0.213,  & 5,6,7,8,9] & \\
&  0.1745, 0.143, 0.117, 0.096]  & & \\
\hline
\end{tabular}
\subcaption*{*Normalized tap coefficients for a sampling duration of $T_s=520.833 \text{ ns}$. \\
$^\dagger$ Velocity values shown for a center frequency of $f_c=2 \text{ GHz}$. For the $700 \text{ MHz}$ band, velocity scales by a factor of $\frac{20}{7}$.}
\end{table}

\begin{figure}[t]
\centering
\includegraphics[width=2.9in]{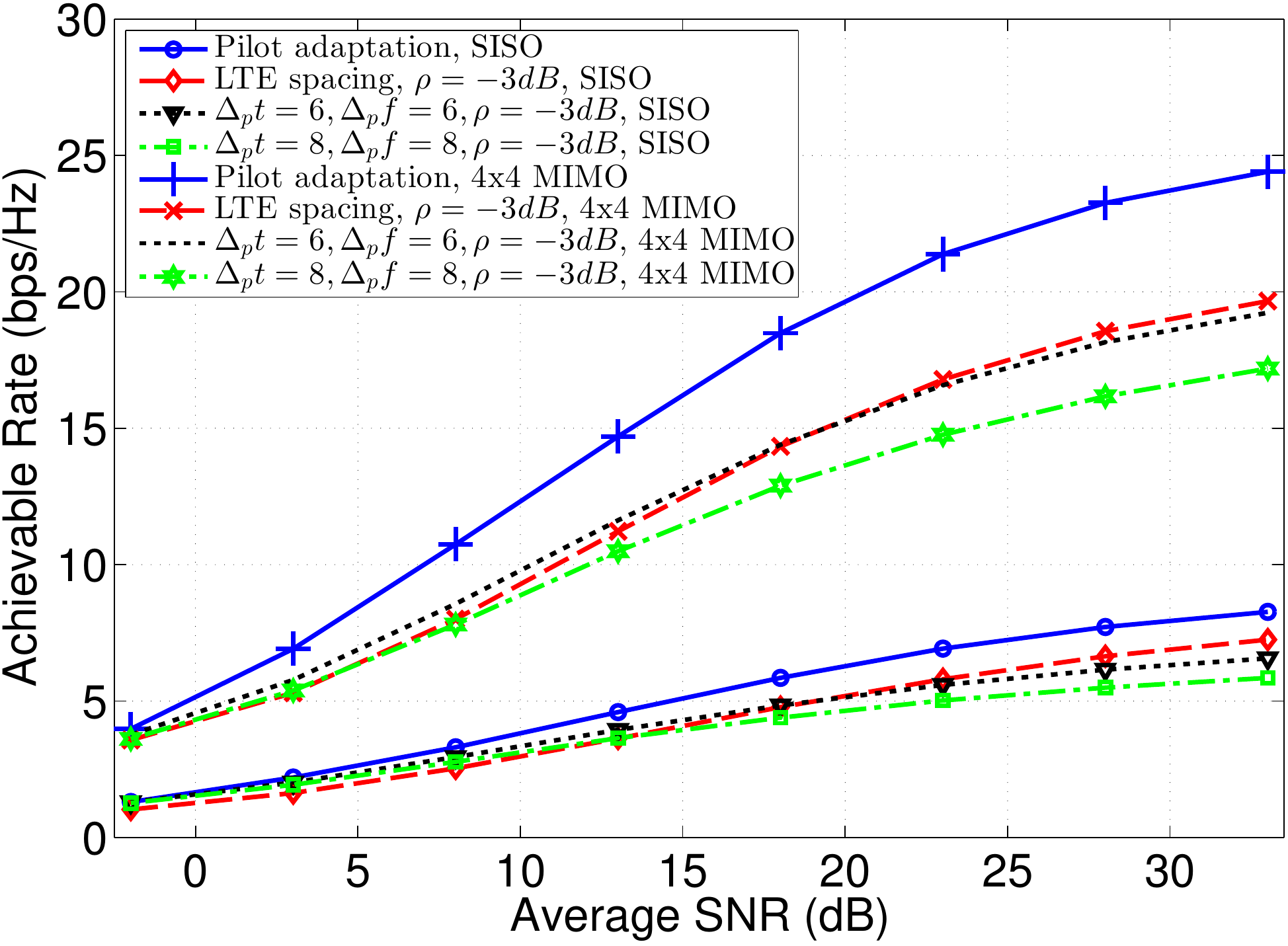}
\caption{Performance of adaptive and pilot schemes for SISO and $4 \times 4$ MIMO-OFDM at $f_c = 2 \text{ GHz}$ in nonstationary UAV to ground/UAV wireless channels.}
\label{Fig8_UAV_SISOMIMO_2GHz_Cap}
\end{figure}

\begin{figure}[t]
\centering
\includegraphics[width=3.0in]{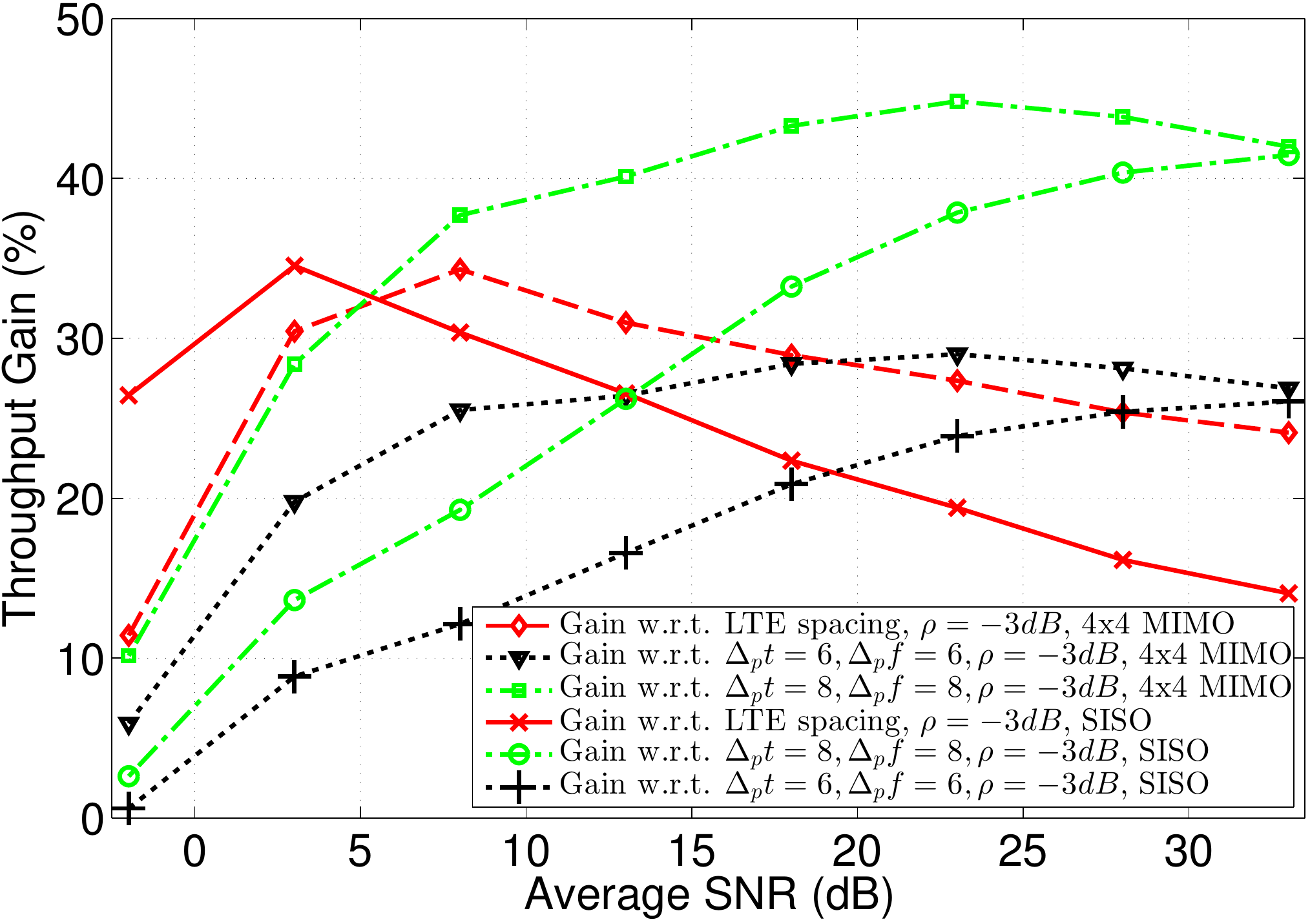}
\caption{Throughput improvement of adaptive pilot configuration over fixed pilot schemes for SISO and $4 \times 4$ MIMO-OFDM at $f_c = 2 \text{ GHz}$ in nonstationary UAV to ground/UAV wireless channels.}
\label{Fig9_UAV_SISOMIMO_2GHz_Gain}
\end{figure}

Fig. \ref{Fig8_UAV_SISOMIMO_2GHz_Cap} shows the achievable rate (throughput) of adaptive pilot configurations for SISO and MIMO with fixed pilot configurations, for UAV to ground/UAV channels at $f_c = 2 \text{ GHz}$. We observe that it outperforms fixed pilot schemes for all SNR values for both SISO and MIMO scenarios. The performance gap increases with SNR, thus showing that the pilot pattern adaptation performs better in low noise (low $\sigma_n^2$) conditions. Fig. \ref{Fig9_UAV_SISOMIMO_2GHz_Gain} shows that using our algorithm, adaptive pilot patterns can achieve up to $35 \%$ more throughput w.r.t. LTE pilot spacing in SISO and $4 \times 4$ MIMO-OFDM modes. Compared to other fixed pilot configurations, this gain can be as high as $45 \%$. 

\begin{table*}[h]
\caption{Comparison with other pilot adaptation schemes in the literature.}
\label{Tab4_Comparison}
\centering 
\begin{tabular}{|c |c |c |c |c |c|}
\hline
\textbf{Reference} & \textbf{Metric} & \textbf{Adaptation} & \textbf{Channel Environment} & \textbf{Throughput} & \textbf{Additional} \\
& \textbf{Maximized} & \textbf{Parameters} & & \textbf{Gain} & \textbf{Details} \\
\hline
Byun et al. & MSE and BER & $\Delta_p t$ & Indoor, Pedestrian and & max. 5.88\%  & Use a look up table \\
\cite{byun2009adaptive} &  & & Vehicular (max. $f_d=\text{83 Hz}$) & & based approach. \\
\hline
Ali et al.  & Throughput & $\Delta_p t$ & Indoor with max. $\tau_{rms} = \text{550 ns}$, & max. 4.58\%  & Use six threshold levels  \\
\cite{ali2008adaptive} &  & & max. $f_d T_s = 1.2 \times 10^{-2}$ & & of Doppler spread. \\
\hline 
Sheng et al.  & Sum rate & $\rho$ & High-speed train & 9-21\%  & Information-theoretic  \\
\cite{sheng2015data} & & &  max. delay spread = $5 \mu s, $& & approach. \\
& & &  max. $f_d = 740 \text{Hz} $& & \\
\hline
Karami et al.  & Mutual & $(\Delta_p t, \Delta_p f, \rho)$ & max. delay spread = $16T_s$, & 14.29 - 42.86\% $^*$ & Derives optimal power \\
\cite{karami2012channel} & information & & max. $f_d = \text{224 Hz} $ & & allocation across all \\
& & & & & OFDM subcarriers. \\
\hline
Simko et al. & Throughput & $\rho$ & ITU Vehicular A & max. 10\% & Results presented \\
\cite{vsimko2012optimal}  & & & & & for LTE with pilot \\
& & & & & power adaptation.\\
\hline
Simko et al. & Throughput & $(\Delta_p t, \Delta_p f, \rho)$& max. $\tau_{rms} = 800 \text{ns}$, & 3 - 80\% $^\dagger$ (SISO) & Gains partially due \\
\cite{simko2013adaptive} & & & max. $f_d = 1200 \text{Hz}$  & max. $8.5 \times ^\dagger$ & to LTE PHY features$^{\ddagger}$ \\
& & & & ($4 \times 4$-MIMO) & \\
\hline 
\textbf{This work} & Throughput & $(\Delta_p t, \Delta_p f, \rho)$ & max. $\tau_{rms} = 1 \mu s$, & 4.33 - 32.24\% (SISO) & Agnostic to LTE  \\
& & & max. $f_d \approx 950 \text{Hz}$  & 4.81 - 40.26\%  & PHY features. Gains \\
& & & doubly selective and & ($4\times 4$-MIMO) &  averaged over SNR \\
& & & nonstationary & & $\text{from } -3 \text{ to } 33 \text{dB}$ in \\
& & & & & nonstationary channels. \\
\hline
\end{tabular} 
\subcaption*{$^*$ Results for $\text{SNR} = -5\text{ dB}$. Rate improvement is negligible for $\text{SNR}>5 \text{ dB}$.\\
$^\dagger$ Because a pilot pattern is associated with a modulation and coding scheme, the throughput gains at high vehicular speeds is much higher. This value is for $f_d = 1200 \text{Hz, SNR}=14 \text{ dB and } \tau_{rms}=400 \text{ns}$ \cite{simko2013adaptive}.\\
$^\ddagger$ Features such as link adaptation, where the modulation order and coding rate is changed based on the channel quality.} 
\end{table*}

\subsection{Comparison with other Pilot Adaptation Schemes}
In this subsection we compare the performance of our pilot adaptation scheme (without carrier aggregation) with other schemes in the literature. To ensure that there is a uniform metric for comparison, we have considered only those works for which the results of data rate improvement with pilot adaptation are available. 

Byun and Natarajan \cite{byun2009adaptive} aim to minimize feedback delays and synchronization mismatch of pilot spacing information in an OFDM system. Since they prioritize channel estimation MSE and BER performance over spectral efficiency, they lose spectral efficiency in some scenarios. In the best case, their approach yields a $5.9\%$ gain in  average spectral efficiency (please refer Fig. 8 of \cite{byun2009adaptive}). 

Ali et al. \cite{ali2008adaptive} adapt the pilot distribution in OFDM-based WLAN according to the variation level of the channel to maximize the throughput. They adapt pilot spacing in time by using six threshold levels for Doppler spread. Their approach performs best in slow-varying channels.

Sheng at al. \cite{sheng2015data} propose to maximize the sum rate using a power allocation scheme between pilot and data symbols for OFDM in a high-speed train (HST) environment. The authors use an information-theoretic approach to solve this problem, by first estimating the average channel complex gains and then using it in a HST basis expansion channel model to formulate a rate-maximization problem.

Karami and Beaulieu \cite{karami2012channel} design a joint adaptive power loading and pilot spacing algorithm to maximize the average mutual information between the input and output of OFDM systems. They derive expressions for the optimal power distribution across all OFDM subcarriers as well. They obtain the best rate improvements in low mobility and low SNR conditions. For high mobility, the throughput improvement reduces significantly. For $\text{SNR}>5 \text{dB}$, there is no noticeable improvement in the rate. 

Simko, Wang and Rupp \cite{vsimko2012optimal} consider optimal power allocation between pilot and data symbols in an OFDM system, and apply it to a LTE system. The authors consider two channel estimation algorithms: Least Squares (LS) and Linear MMSE (LMMSE). The best case throughput improvement is reported to be 10\%. 

Simko et al. \cite{simko2013adaptive} consider joint optimization of pilot spacing and power for SISO and MIMO-OFDM systems (without carrier aggregation). Like in \cite{vsimko2012optimal}, they compare the throughput of their adaptation and power allocation algorithm with that of a standard LTE system. They propose mapping the pilot pattern to the modulation and coding scheme (MCS) of LTE. The combination of (a) change in modulation order and code rate and (b) change in pilot power and spacing, can result in very high gains (upto $8.5\times$) at very high vehicular speeds with $4\times 4$-MIMO.

In contrast to the above, our results are agnostic to the LTE protocol, and hence applies to any general OFDM-based system. However, we do compare the throughput performance of our scheme with Normal Cyclic Prefix (CP) LTE \cite{sesia2011lte} for vehicular and air-to-ground wireless channels. For both channels, the performance gains w.r.t. LTE range from 16.68-27.49\% (refer Table \ref{Tab5_cap_gains}). Unlike the other works mentioned above, our results are averaged for a nonstationary channel scenario of a slow fading line-of-sight channel that evolves to a fast fading multipath channel. Our results demonstrate that even when channel statistics gradually changes in a timescale of a few hundred milliseconds (in our case, $\approx 108 \text{ms}$), our proposed scheme gives a significant throughput improvement which has not been reported before, to the best of our knowledge. The summary of the key results from the above works is summarized in Table \ref{Tab4_Comparison}.

\subsection{Pilot Adaptation in Multi-band CA-OFDM Systems}
We simulated adaptive pilot configurations for multi-band carrier aggregation OFDM (CA-OFDM) systems, for the nonstationary channel scenarios shown in Fig. \ref{Fig7_channel_stats_varn_all} for $N_b=2, f_1 = 700 \text{ MHz and }f_2 = 2 \text{ GHz}$. We use the channel profile codebook $\mathcal{R}_C$ shown in Table \ref{Tab6_codebook_profiles} for both frequency bands. 
We compare the throughput results for CA SISO and MIMO-OFDM systems in (a) UAV to ground/UAV channels and (b) terrestrial channels. 

\begin{figure}[t]
\centering
\includegraphics[width=3.0in]{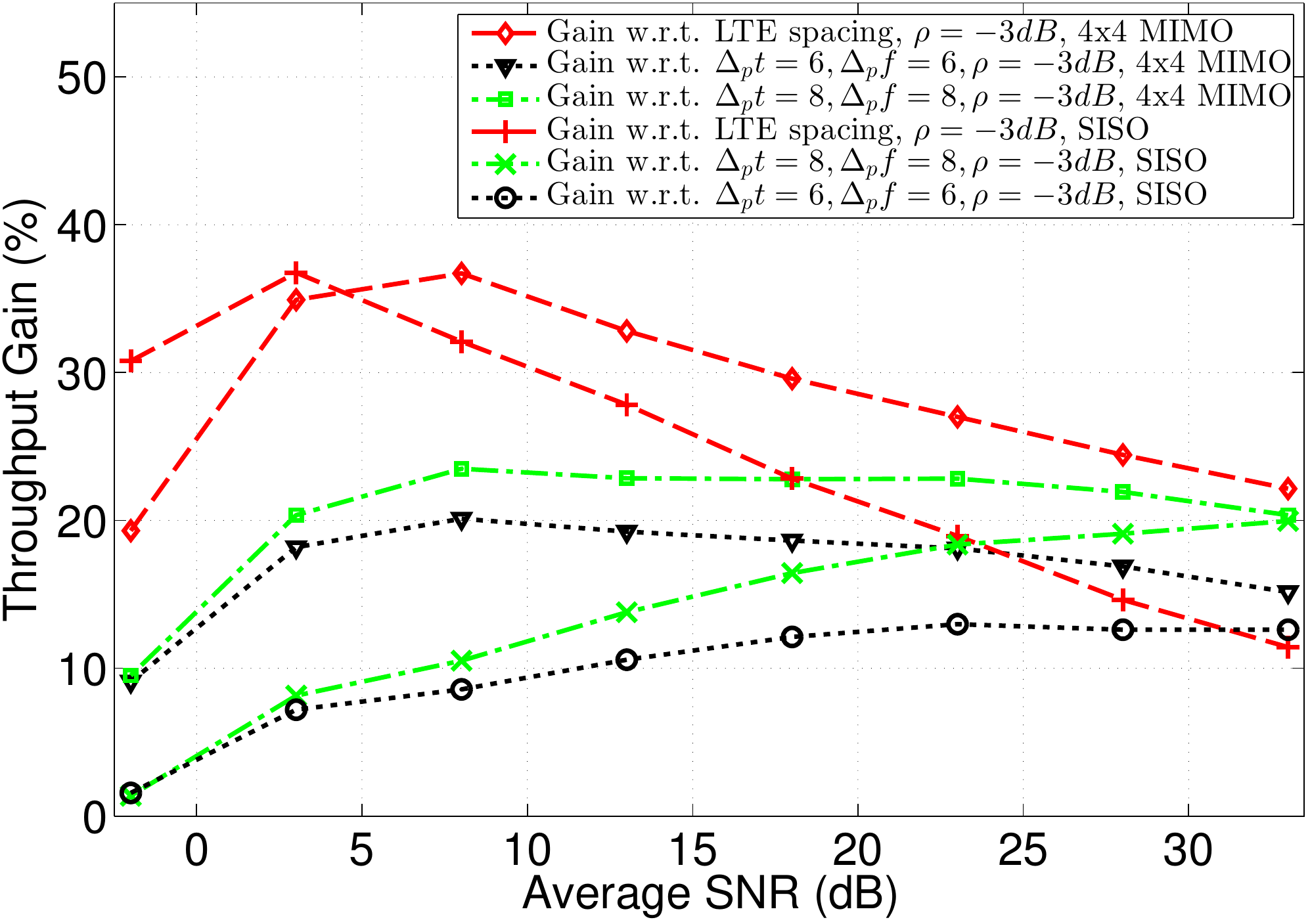}
\caption{Throughput improvement of adaptive pilot configuration over fixed pilot schemes for a multi-band CA-OFDM system with component carriers at $f_1 = 700 \text{ MHz and } f_2 = 2 \text{ GHz}$, in nonstationary UAV to ground/UAV wireless channels.}
\label{Fig11_UAV_SISOMIMO_CarrAgg_Gain}
\end{figure}

\begin{figure}[t]
\centering
\includegraphics[width=3.0in]{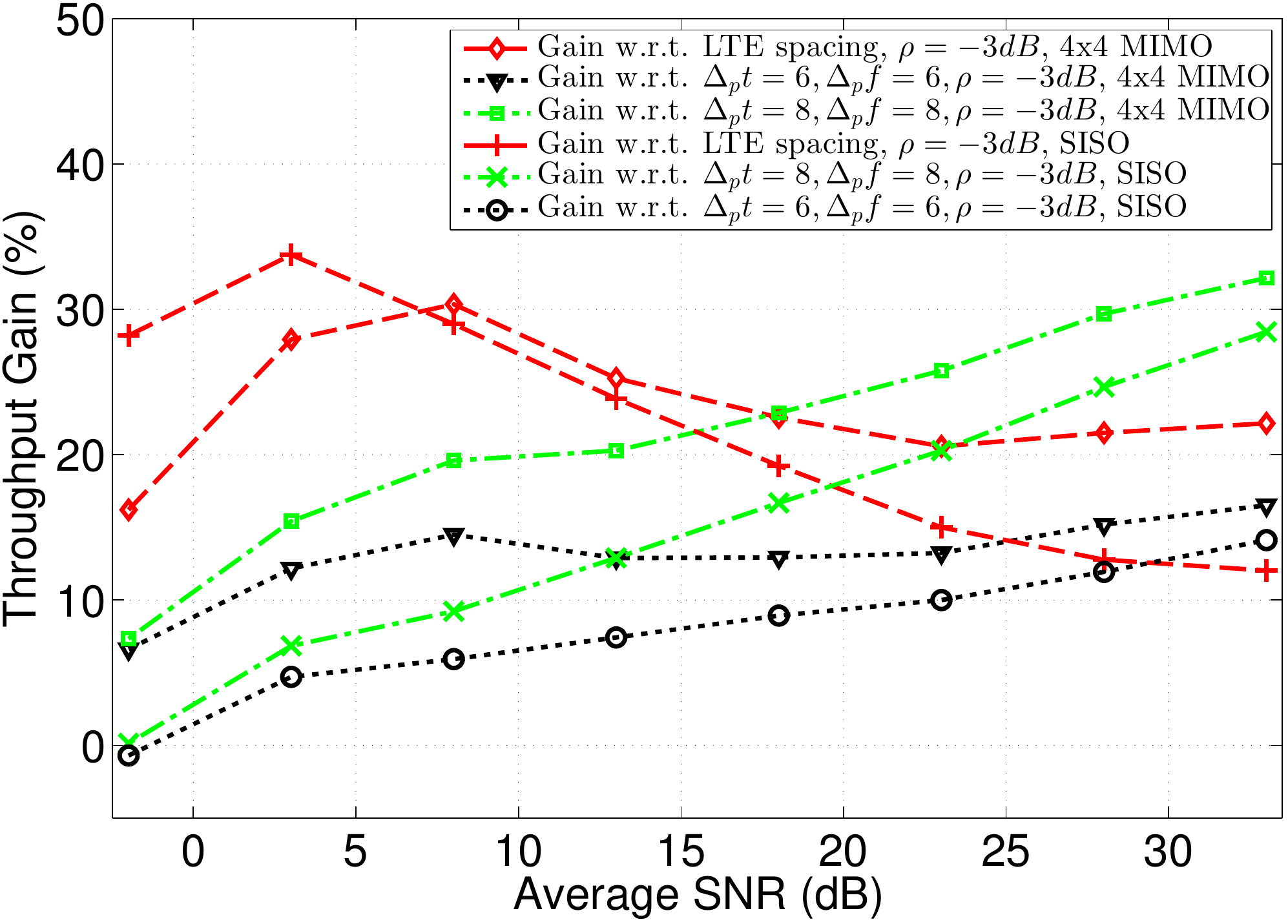}
\caption{Throughput improvement of adaptive pilot configuration over fixed pilot schemes for a multi-band CA-OFDM system with component carriers at $f_1 = 700 \text{ MHz and } f_2 = 2 \text{ GHz}$, in nonstationary terrestrial wireless channels.}
\label{Fig12_Terr_SISOMIMO_CarrAgg_Gain}
\end{figure}

Figures \ref{Fig11_UAV_SISOMIMO_CarrAgg_Gain} and \ref{Fig12_Terr_SISOMIMO_CarrAgg_Gain} show the throughput gains achieved by adaptive pilot configurations w.r.t. the fixed pilot configurations considered for UAV to ground/UAV and terrestrial wireless channels respectively. The gain is generally higher for UAV wireless channels as compared to terrestrial systems. This is so because $\Delta_p f$ can be increased to improve the spectral efficiency of the typically frequency-flat air-to-ground/air-to-air wireless channels.

\begin{table*}[t]
\caption{Average throughput gain (in \%) of adaptive pilot configuration compared to fixed pilot configurations}
\label{Tab5_cap_gains}
\centering 
\begin{tabular}{c c c c c c c c c c c c c}
\hline
 & \multicolumn{6}{c}{Terrestrial Channels} & \multicolumn{6}{c} {UAV to ground/UAV channels} \\
 Pilot Spacing & \multicolumn{2}{c}{$f_c = 700$ MHz} & \multicolumn{2}{c}{$f_c = 2$ GHz} & \multicolumn{2}{c}{CA} & \multicolumn{2}{c}{$f_c = 700$ MHz} & \multicolumn{2}{c}{$f_c = 2$ GHz} & \multicolumn{2}{c}{CA} \\
($\rho=-3\text{ dB}$) & SISO & MIMO & SISO & MIMO & SISO & MIMO & SISO & MIMO & SISO & MIMO & SISO & MIMO \\
\hline
$\Delta_p f = \Delta_p t = 6$ & 4.33 & 10.96 & 20.44 & 26.30 & 11.35 & 17.31 & 4.32 & 9.46 & 16.53 & 20.18 & 9.69 & 13.97 \\
$\Delta_p f =\Delta_p t = 8$ & 4.81 & 9.82 & 32.24 & 40.26 & 16.17 & 21.59 & 8.99 & 14.94 & 31.98 & 39.62 & 18.64 & 24.72 \\
LTE normal CP & 19.36 & 27.49 & 20.58  & 27.13 & 19.93 & 27.33 & 18.83 & 24.84 & 16.68 & 20.60 & 17.82 & 22.92 \\
\hline
\end{tabular} 
\end{table*}

\begin{figure}[t]
\centering
\includegraphics[width=3.0in]{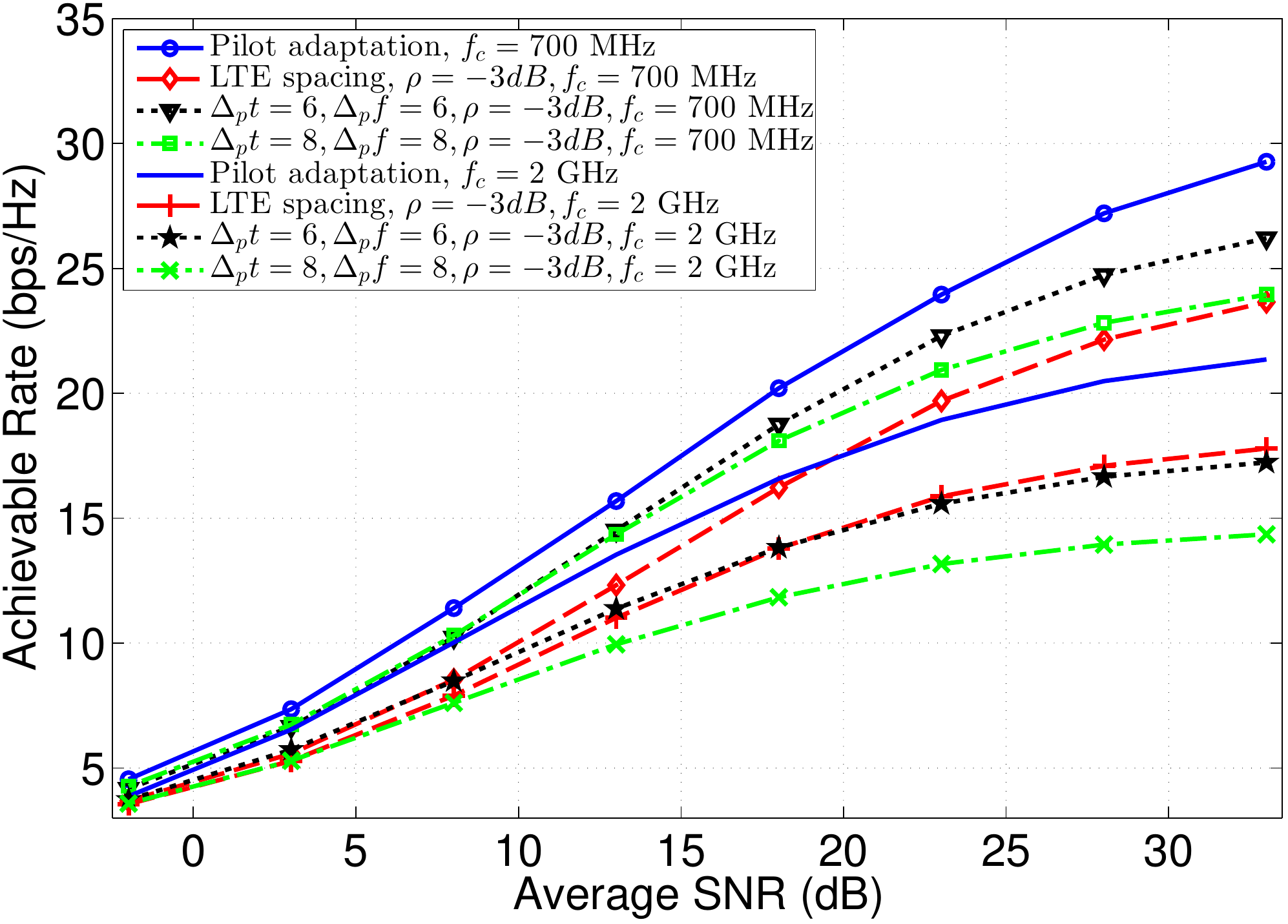}
\caption{Enhancement of the achievable rate at $f_1 = 700 \text{ MHz}$ compared to $f_2 = 2 \text{ GHz}$, for a $4 \times 4$ MIMO-OFDM system in nonstationary terrestrial wireless channels.}
\label{Fig13_Terr_MIMO_700MHz2GHz_Cap}
\end{figure}

Similarly, the channel in the $700 \text{ MHz}$ band will tend to be more benign in terms of temporal fading, due to the relatively low Doppler spread when compared to that at $f_2 = 2 \text{ GHz}$. Hence, in this case $\Delta_p t$ of the component carrier in the $700 \text{ MHz}$ band can be increased w.r.t. that in the $2 \text{ GHz}$ band. Fig. \ref{Fig13_Terr_MIMO_700MHz2GHz_Cap} shows the relative performance enhancement in the $700 \text{ MHz}$ band, validating the above. Hence, adapting the pilot density in two or more different operating frequency bands presents a means to increase the channel capacity, and offers an additional degree of freedom for cross-layer optimization and load-balancing algorithms in CA-OFDM systems.

Table \ref{Tab5_cap_gains} summarizes the average throughput gain achieved by pilot adaptation (averaged over SNR) w.r.t. the fixed pilot schemes considered in this section. It shows that adaptive pilot configurations provide an average throughput (achievable rate) gain of ${\sim}20\%$ when compared to current LTE systems, with peak capacity improvements of $35\%$. 

This enhancement comes without noticeably increasing the computational complexity, or the communication overhead between the transmitter and the receiver. Typical MMSE receivers rely on estimated second order channel statistics to enhance performance \cite{Arslan_OFDM_chanest_survey_2007}, and turbo decoders rely on estimated noise power to compute the log-likelihood ratios (LLRs). The signaling involved for the codebooks of Table Table \ref{Tab6_codebook_profiles} is negligible: up to $\lceil \log_2 (M_t M_f) \rceil = \lceil \log_2 (6 \times 4) \rceil = 5\text{ bits}$ are required once in every $T_{ofdm} = 1500$ OFDM symbols ($107.8 \text{ ms}$ for the typical $15 \text{ kHz}$ subcarrier spacing). For CA-OFDM with $N_b = 2$ for our example, $\lceil \log_2 [M_t M_f + (N_b - 1) M_f] \rceil = \lceil \log_2 (6 \times 4 + 4) \rceil = 5\text{ bits}$ are necessary with our reduced feedback scheme, as compared to $\lceil \log_2 (N_b M_t M_f) \rceil = \lceil \log_2 (2 \times 6 \times 4) \rceil = 6\text{ bits}$ that would've been necessary otherwise. In both cases (with and without CA), the feedback of codebook indices translates to a data rate overhead of $46.38 \text{ bps}$, which is negligible compared to the peak data rates achieved by current OFDM-based wireless standards.
%

\section{Practical Considerations}\label{Pract_consid}
For pilot adaptation in the downlink frame, the proposed scheme relies on channel statistics estimated by the user that are fed back to the base station. Hence, the pilot patterns can vary among users in a cell because different users generally experience different channel statistics. This implies that implementing pilot adaptation is not straightforward for pilots that are broadcasted in a cell, such as the Cell-Specific Reference Signal (CRS) in LTE \cite{sesia2011lte}. 

Moreover, there are possibilities of pilot corruption due to pilot contamination between two cells if all types of pilots are adapted. Hence, we provide a few guidelines for pilot pattern adaptation:
\begin{enumerate}
\item It is well suited for user-specific pilots, for example UE specific Reference Signals of LTE \cite{sesia2011lte}.  

\item It is also applicable for peer-to-peer links such as wireless backhaul, vehicular-to-vehicular, UAV-to-ground/UAV-to-UAV systems. In such systems, the issue of interference with other pilots typically does not arise. 

\item It can be extended to grouping of users having similar channel conditions during resource allocation \cite{Ksairi_5G_MUMIMO_PilAdapt_2016}. Active user-aware dynamic pilot distribution as well as joint pilot and user scheduling can help better utilize maximize system spectral efficiency. 
\item It is synergistic with CA-OFDM; users can be grouped and scheduled to different component carriers according to their channel statistics.

\item 
In multiuser-MIMO (MU-MIMO), scheduling of users with (a) similar second order channel statistics and (b) orthogonal precoding vectors is necessary to perform pilot pattern adaptation. 
\end{enumerate}

\section{Conclusion}\label{conc}
Flexibility is a key trait of future wireless standards, where the communications protocols can be customized on a per-user or network basis to optimize performance. In this paper, we provided an efficient heuristic algorithm to design rate maximizing pilot configurations in SISO and MIMO-OFDM systems based on the second order statistics of doubly selective nonstationaty wireless channels. We also extended this concept to CA-OFDM systems. We derived closed form expressions for channel estimation MSE for pilots arranged in a ``diamond-pattern''. Using the derived MSE expressions and the lower bound on inter-carrier interference (ICI), we devised a codebook-based approach to adapt the pilot spacing and power based on estimated channel statistics. Our scheme adds negligible computational complexity since (a) modern wireless receivers implementing the MMSE receiver already implement such channel statistics estimators, and (b) finding the closest codebook profile and the optimal pilot configuration are also low-complexity operations. Also, the feedback overhead is shown to be negligible in current, high-capacity wireless standards. 

Our numerical results for two communications environments have shown that the average throughput gain of our scheme w.r.t. LTE pilot spacing is $16 \text{ to } 20 \%$ for SISO and $20 \text{ to }28\%$ for $4 \times 4$ MIMO. Our algorithm is agnostic to standard-specific metrics mechanisms such as adaptive modulation and coding. Therefore, the presented results are fundamental and representative of throughput gain by using adaptive pilot configurations in OFDM-based wireless standards.

Adaptive waveforms is the theme of physical layer design of future 5G wireless communication systems in order to maximize the spectral efficiency. 
Extending this work to other multicarrier waveforms such as Filter-bank Multicarrier (FBMC) and Non-orthogonal multiple access (NOMA) schemes would be a meaningful contribution towards enhancing the spectral efficiency of other candidate physical layers for 5G.

Pertaining to cross-layer optimization using adaptive pilot configurations, open research areas include joint pilot design and user scheduling in carrier-aggregation and MU-MIMO wireless systems. Protocol designs built around this framework which drive the selection of other radio resource management (RRM) parameters, and characterization of their performance would be crucial to evaluating the enhancement in overall network throughput of such systems.



\bibliographystyle{IEEEtran}
\bibliography{references_tvt}
\ifCLASSOPTIONcaptionsoff
  \newpage
\fi

\end{document}